\pdfoutput=0
\documentclass[useAMS,usenatbib]{mn2e}
\usepackage{graphicx,bm,mathabx,braket,dsfont}

\title[Approximate Bayesian Computation for Astronomical Model
Analysis]{Approximate Bayesian Computation for Astronomical Model
  Analysis: A Case Study in Galaxy Demographics and Morphological
  Transformation at High Redshift}
\author[E. Cameron and A. N. Pettitt]{E. Cameron$^{1}$\thanks{E-mail:
dr.ewan.cameron@gmail.com} and A. N. Pettitt$^{1}$\\
$^{1}$School of Mathematical
Sciences (Statistical Science), Queensland University of Technology
(QUT), GPO Box 2434,\\ Brisbane 4001, QLD, Australia}
\begin{document}

\date{Submitted to MNRAS: 7 Feb 2012.}

\pagerange{\pageref{firstpage}--\pageref{lastpage}} \pubyear{2011}

\maketitle

\label{firstpage}

\begin{abstract}
``Approximate Bayesian Computation'' (ABC) represents a powerful
methodology for the analysis of complex stochastic systems
for which the likelihood of the observed
data under an arbitrary set of input parameters may be entirely \textit{intractable}---the latter condition rendering useless the
standard machinery of tractable likelihood-based, Bayesian statistical inference (e.g.\
conventional Markov Chain Monte Carlo simulation; MCMC).  In this article we demonstrate the potential of ABC for astronomical model analysis by application to a
case study in the morphological transformation of high
redshift galaxies.  To this end we develop, first, a stochastic model
for the competing processes of merging and secular evolution in the
early Universe; and second, through an ABC-based comparison against the observed
demographics of massive ($M_\mathrm{gal}>10^{11}$$M_\odot$) galaxies (at
$1.5 < z < 3$) in the CANDELS/EGS dataset we derive posterior probability densities
for the key parameters of this model.  The ``Sequential Monte Carlo''
(SMC) implementation of ABC exhibited herein, featuring both a
self-generating target sequence and self-refining MCMC kernel, is amongst the most efficient of contemporary approaches
to this important statistical algorithm.  We highlight as well through our chosen case study the value of careful
summary statistic selection, and demonstrate two modern strategies
for assessment and optimisation in this regard.  Ultimately, our ABC
analysis of the high redshift morphological mix returns tight
constraints on the evolving merger rate in the early Universe and
favours major merging (with disc survival or rapid reformation) over
secular evolution as the mechanism most responsible for building up the first generation of bulges in early-type disks.
\end{abstract}


\section{Introduction}\label{introduction}
With origins in population genetics and evolutionary
biology (e.g.\ \citealt{tav97,pri99,bea02}; and see \citealt{csi10} for a
recent review) Approximate Bayesian Computation (ABC) offers a
powerful technique for recovering posterior probability densities
from complex stochastic models for which the likelihood may be
entirely \textit{intractable}.  That is, \textit{the probability of the
observed data under a given set of input parameters cannot be solved
analytically or computed directly (within a practical timeframe).}  Examples include the estimation of time to the most
recent common ancestor under the coalescent model with
recombination given a full suite of modern DNA sequencing \citep{mar06}, or the derivation of transition probabilities
in continuous time Markov models of macroparasite population
evolution from simple demographics \citep{dro10}.  However, although
there exist a variety of important \textit{astrophysical} models
with inherently intractable likelihoods (a number of which we will
discuss herein), applications to-date
of ABC in this field remain surprisingly rare.\footnote{Indeed the
  authors can find no astronomical reference to either the terms ``approximate
  Bayesian computation'' or ``likelihood-free'' (inference) on the
  NASA ADS database; and Google Scholar indicates no astronomical
  citations yet to any of the biological/mathematical ABC literature
  mentioned herein.  A more pedagogical treatise on the potential for
  ABC in astronomy presented by Chad Schafer and Peter Freeman 
at the Statistical Challenges in Modern Astronomy V conference in 2011
(available at
URL[http://www.springer.com/statistics/book/978-1-4614-3519-8/]),
which 
details two interesting uses for ABC in extra-galactic data analysis,
represents to our knowledge the only prior application in this field.}  The only indispensable ingredients required for ABC are: \textsc{(i)} a \textit{stochastic model} for the observed
data, replicating the behaviour of all random processes driving the system at hand, as well as any relevant observational errors; and \textsc{(ii)} a
\textit{discrepancy measure}, based typically on a set of low-order summary statistics,
to quantitatively gauge similarity between output from this model and the
empirical benchmark.

One potentially valuable role for ABC in an
astronomical context \textit{may} thus be in the constraint of semi-analytic
models (SAMs) of galaxy formation (cf.\
\citealt{col00,ben03,bau06,bow06,del10,nei10})---in which the output at run-time necessarily exhibits
complex stochasticity owing to the effects of cosmic variance
(induced computationally via sampling from within large-scale dark matter
simulations, \citealt{spr01,kne11}; or via Monte Carlo construction of halo
merger trees, \citealt{lac93,par08}).  For ABC analysis of such codes
an appropriate discrepancy measure might then be the metric
distance between simulated and observed luminosity functions
under a sensible binning scheme.\footnote{For readers familiar
  with the work of \citet{bow10} we note that the ``discrepancy
  parameter'' introduced for their emulation of the GALFORM SAM could
  not be employed as such in ABC as it is not (designed as) a gauge of model--data
similarity; indeed it serves an entirely different purpose in their
analysis, acting as an error term for cosmic variance and structural
uncertainty in their code.}  With conditions (\textsc{i}) and
(\textsc{ii}) above thus satisfied ABC offers an easily-implemented,
theoretically well-established \citep{nun10,mar11,fea12} alternative
to the computationally intensive ``approximate likelihood'' approach
(requiring very large scale simulation/re-simulation, e.g.\
\citealt{woo10,hen09,lu11}; and note \citealt{ben11} regarding the required
diversity of merger trees sampled
for genuine convergence of SAMs), or the user-intensive application of
model emulators (requiring a non-trivial degree of run-time supervision
and operator expertise, cf.\
\citealt{bow10} and references therein).\footnote{As a caveat to the above referencing we
  note that: \textsc{(i)} though the analyses of \citet{hen09} and \citet{lu11}
  are both conducted broadly in the style of the ``approximate
  likelihood'' approach formalised by \citet{woo10} there are also a
  number of significant implementational differences unique to each; and \textsc{(ii)} though the work of \citet{kam08} has in
  previous papers been cited as an example of MCMC-based SAM
  constraint, in fact, their study concerns a purely analytic model
  for which there exists no intrinsic stochasticity (thus, only
  approximate observational errors enter their likelihood
  computation).  Finally, we
refer the interested reader to \citet{har11} for a concise overview of
the similarities and differences between the ``approximate
likelihood'' and ABC approaches to
inference from statistical simulation, and to \citet{not11} for an
advanced treatment of the link between a particular version of ABC and
the Bayes Linear technique
(cf.\ \citealt{gol07}) underlying the model emulator approach.}

Another astronomical problem readily amenable to ABC is
that of inferring the age and mass of an unresolved star cluster based
on its broadband spectral energy distribution (SED).  Here it is the
sheer diversity/complexity of evolutionary tracks 
open to a cluster of given mass under a
stochastically sampled initial mass function (IMF) that renders
unfeasable (i.e., intractable) any explicit formulation of the observational likelihood
function (cf.\ \citealt{asa12,bon12,her12,kod12})---though with brute-force re-simulation at fixed input using a cluster
formation code such as SLUG
\citep{fum11,das11} or MASSCLEAN \citep{pop09} one can in principle
generate a fair approximation to it by recording the frequency of
output in each region of the observational hyperspace.  Indeed with huge
libraries of such simulations \citet{pop10} and \citet{fou10} are
already employing this approximate likelihood approach
for ``first-order'' cluster mass and age estimation.  An appreciation
of the established ABC method may offer
practioners in this field valuable insight
into the challenges they face, which are, in abstraction, already addressed routinely
 in the related statistical literature.  For instance, the merits of alternative
 filter combinations may be readily assessed through the lens of
 \textit{summary statistic selection}, and a realistic \textit{distribution} of cluster metallicities
and dust reddening vectors robustly accounted for via the Bayesian
technique of \textit{marginalising over nuisance parameters}.

Another two intriguing examples of astronomical model analysis
problems amenable to ABC appear in recent work by \citet{hek11} and
\citet{lei12} in the disparate fields of asteroseismology and IMF
profiling, respectively.  In the former it is the non-linear
propogation of realisation noise in the solar oscillation spectrum
that renders intractable the observational likelihood function.
Simulated datasets though may be readily generated for this system, and
\citet{hek11} have identified a corresponding set of summary statistics optimal for
inference of the key model parameters.  Specification of an
appropriate discrepancy distance thus remains the final (and relatively trivial)
hurdle to ABC implementation here.  In the \citet{lei12} study
it is the intrinsic complexity of two-body relaxation 
within many-body stellar systems that necessitates a simulation-based
approach to likelihood approximation.  The cluster metallicity and the global
binary fraction act as nuisance parameters of their model, while binary star
confusion and the (inherent) projection of a 3D system onto the 2D
observational plane contribute complex 
sources of measurement ``error'' best treated by forward simulation.

In this paper we illustrate heuristically the power of ABC for astronomical model
analysis through application to yet another branch of this rich subject, namely the morphological transformation
of massive galaxies at high redshift.  In particular, we demonstrate a
contemporary Sequential Monte Carlo (SMC) formulation of the ABC algorithm (cf.\
\citealt{del06,sis07,dro10}), as well as a
regression-based procedure for constructing an optimal summary
statistic--discrepancy measure pairing for the purpose of parameter estimation
\citep{fea12}.  Importantly, the stochastic model we explore herein
features both an ``independent evolution'' case for which the
likelihood is in fact tractable and a ``co-evolution'' case
for which it is not---the former allowing the strengths and limitations of our ABC-based
solution to be established against conventional Markov Chain Monte
Carlo (MCMC) simulation and the latter a demonstration of the unique
possibilities of ABC analysis.

Installation of the new Wide-Field Camera 3 (WFC3) on the
Hubble Space Telescope (HST) in 2009---and the subsequent allocation of
vast amounts of observing time to deep, near-infrared (NIR) surveys with this
instrument, including the Early Release Science program (ERS; \citealt{win11}) and the
Cosmic Assembly Near-IR Deep Extragalatic Legacy Survey (CANDELS;
\citealt{gro11,koe11})---has at last made accessible (at high
resolution) the rest-frame optical morphologies of distant
galaxies at the epoch of peak cosmic star formation and AGN activity
($z\sim2$; \citealt{lil96,mad96,oes12,war94}).  Early studies exploiting
these new datasets have documented the emergence of the first Hubble sequence
analogues \citep{cam11,con11a,szo11a}, demonstrated the compactness of the first massive
spheroids \citep{szo10,szo11b,new11}, explored the unique characteristics of
 galaxies
ultraluminous at infrared \citep{kar11} and X-ray wavelengths
\citep{koc11,ros11,sch11}, and probed structural transformation in extreme cluster
environments \citep{lot11,pap11}.  Thus far, however, there have been
remarkably few studies to exploit the full potential of \textit{demographic analysis}
for constraining pathways of galaxy evolution---one early exemplar being
Bell et al.'s (2011) search for correlations between the global observables
of key galaxy sub-populations in the CANDELS dataset divided coarsely by rest-frame optical morphological type (via
the usual proxy of global S\'ersic index; cf.\
\citealt{dri06,cam09,kelv11}).  Hence, we have chosen here specifically for our exposition
of the ABC technique a
case study in the demographic analysis of WFC3 data in the hope of motivating further research in
this direction.

The structure of this paper is as follows.  In Section \ref{dataset}
we review the publicly-available source catalogues and images
comprising our high redshift demographic benchmark, then in Section
\ref{method} we present the core of our case study in ABC for
astronomical model analysis.  First, we describe our model for galaxy evolution and our procedure for stochastic simulation from this model (Section \ref{stochastic}).
Second, we explain the ABC algorithm and the SMC
approach to its implementation (Section
\ref{smcabcalg}).  Third, we examine in depth the important process of
constructing an optimal summary
statistic--discrepancy parameter pairing (Section \ref{sumstats}).
And fourth,
we confirm the general similarity between our ABC
and MCMC posteriors in the tractable ``independent evolution'' case, and present our final ABC-only posteriors for the more
realistic, but likelihood intractable, ``co-evolution'' case (Section
\ref{smc}).  In Section \ref{results} we conclude this paper with a
discussion of the
implications of the model
constraints so derived for astrophysical theories of
morphological transformation in the early Universe.

We have thus attempted to organise our exposition of ABC in such a
manner as to allow astronomers interested in this important
statistical algorithm but not
working directly in the area of galaxy evolution to optionally skip over
the technical details and justification of our model (Section
\ref{stochastic}) without disadvantage (instead
reading only Sections \ref{smcabcalg}, \ref{sumstats}, and \ref{smc}
in depth).  All magnitudes are quoted in the AB system and
a standard $\{ \Omega_M = 0.3, \Omega_\Lambda=0.7, h=0.7\}$
cosmological model is adopted throughout.

\section{Data}\label{dataset}
Featuring a vast ensemble of multiwavelength imaging compiled from
both ground-based and space-based observatories the Extended Groth Strip (EGS) region of the Northern Sky (centred on
RA: $14^{h}17^{m}$, Dec: $+52^{\circ}30^{\prime}$) numbers amongst the premier legacy survey fields of the
modern era. The All-wavelength Extended Groth strip International Survey team (AEGIS;
\citealt{dav07}) has been responsible for the bulk of this data
collection through extensive observational campaigns with HST and
Spitzer.  Such a comprehensive set of photometric measurements greatly facilitates the
estimation of redshifts and stellar masses via
SED template fitting, and there exist a number of published
studies characterising the high redshift galaxy population in the EGS to this effect.

\begin{figure*}
\hspace{-0.85cm}\includegraphics[width=11.25cm]{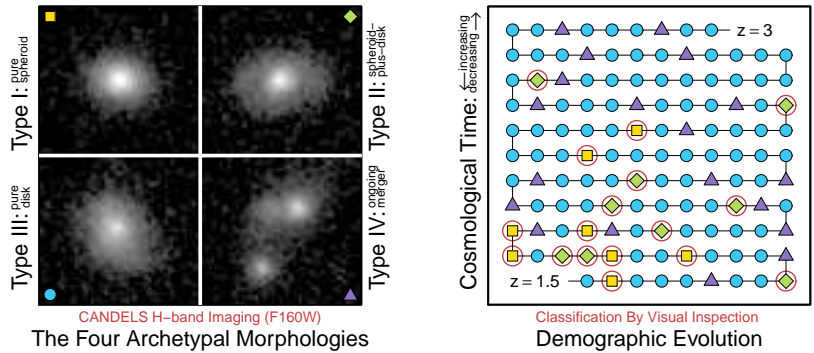}
\caption{(\textit{Left:}) CANDELS (HST WFC3/IR) $H$-band postage stamp images
  characterising the four archetypal morphologies present amongst massive
  ($M_\mathrm{gal} > 10^{11}$$M_\odot$), high redshift ($1.5 < z < 3$) galaxies
  in the B11 dataset.  (\textit{Right:}) An illustration of
  demographic evolution (i.e., \textit{the evolving morphological
  mix}) amongst our B11 (CANDELS/EGS) sample.  The symbol key is given
  in the lefthand panel, and the ``evolved'' morphological types
  (pure
  spheroids and spheroid-plus-disks) are circled in red to highlight
  their late build up.\label{morphologies}}
\end{figure*}

In this study we employ the publicly-available and up-to-date,
ultra-violet-(UV)-to-far-infrared-(FIR)-based catalogue of \citet{bar11a,bar11b} [B11
hereafter] to identify a complete sample of high mass ($M_\mathrm{gal} >
10^{11}$$M_\odot$), early Universe ($1.5 < z < 3$) systems.  The B11
photometric redshifts, based on up to 19 band flux measurements in the
survey core, feature an overall accuracy (measured against
a spectroscopic subsample from AGEIS with median $z \sim 1.3$) of $\Delta z / (1+z) =
0.034$ at a sub-2\% catastrophic failure rate.  At the highest
redshifts ($z > 2.5$) comparison against a spectroscopic sample of 91
Lyman-Break Galaxies (LBGs) confirms only a slight degradation to
$\Delta z / (1+z) = 0.069$.  The corresponding B11
stellar masses for these systems were derived using the PEGASE\footnote{PEGASE: Projet d'Etude
  des GAlaxies par Synthese Evolutive \citep{fio97}.} SED library (with Salpeter
initial mass function and Calzetti extinction)---the choice of which (from amongst the
wide range of alternative SED libraries) represents the
dominant source of systematic uncertainty here (of order
0.1-0.3 dex; \citealt{bar11b}).  For the purposes of this paper, in
which our principle aim is to demonstrate as straightforwardly as possible the
technicalities of the ABC approach, we hereafter neglect further
quantitative consideration of
these uncertainties (except when required for fitting the build up in number
density over cosmic time, which contributes two nuisance parameters to our model, in Section
\ref{stochastic}).

The CANDELS team \citep{gro11,koe11} is currently engaged in the
acquisition of high-resolution,
near-infrared (and UV) imaging targeting distant galaxies in selected
sub-regions of five key legacy fields (GOODS-N, GOODS-S\footnote{GOODS-N/-S: Great Observatories Origins Deep -North/-South.}, the
EGS, COSMOS\footnote{COSMOS: COSMic evOlution Survey.}, and the
UDS\footnote{UDS: UKIDSS (UKIRT [United Kingdom InfraRed Telescope]
  Infrared Deep Sky Survey) ultra-Deep Survey.}) totaling $\sim$800 arcmin$^2$ under an allocation of 902
orbits of HST/WFC3
exposure time.  Drizzled to a
pixel scale of 0.06 arcsec, the presently-available epoch (egs01) of
imaging within (an $\sim$90 arcmin$^{2}$ sub-region of) the EGS field features a point spread function (PSF) full width half maximum (FWHM) of $\sim$0.18 arcsec and a 5$\sigma$ detection
limit of 26.8 mag in the F160W ($H$-band) filter (with comparable
coverage in the F125W filter).  As such CANDELS already represents the
highest quality dataset published to-date for the study of rest-frame
optical morphologies at $z\sim2$-3 in the EGS.  Accordingly for the
present analysis we derive
our high redshift demographic benchmark from visual classification of all 126 members of the B11 catalogue
at $M_\mathrm{gal}>10^{11}$$M_\odot$ and $1.5 < z < 3$ imaged thus far.

To this end one of us (EC) inspected
each source carefully in the CANDELS (HST WFC3/IR) $H$-band mosaic with
\texttt{ds9} and assigned it one of the following four types: \textsc{(i)}
\textit{spheroid} (compact elliptical; cf.\ \citealt{szo11b}), \textsc{(ii)}
\textit{spheroid-plus-disk} (early-type disk with a prominent
central bulge; cf.\ \citealt{cam11}), \textsc{(iii)}
\textit{pure disk} (late-type, bulgeless [perhaps clumpy] disk; cf.\
\citealt{elm07a}), or \textsc{(iv)} \textit{ongoing merger} (evident
violent relaxation event in progress, as revealed by the presence of distinctive tidal
features and/or multiple massive
nuclei; cf.\ \citealt{elm07b}).  In total we count 8 Type
\textsc{i} spheroids, 9 Type \textsc{ii} spheroid-plus-disks, 90 Type
\textsc{iii} pure disks, and 19 Type \textsc{iv} ongoing mergers in our sample.  Example $H$-band
postage stamp images characterising these archetypal high redshift
morphologies are presented in the lefthand panel of Figure
\ref{morphologies}, as is an illustration of the
demographic evolution across our sample in the
righthand panel.  Once again in accordance with the expository aims of
this paper regarding ABC we do not explore the (complex) possible impacts of classification subjectivity
on our results---although we note that both ABC and the Bayesian
framework in general offer a powerful statistical basis for
marginalising over such sources of uncertainty (cf.\ \citealt{gel03};
E.N.\ Taylor, in prep.; and see our treatment of various nuisance
parameters in Sections \ref{stochastic} and \ref{smcabcalg}), particularly where the experimental
evaluation of the classification system has been appropriately designed and
implemented \citep{han97}.  Reassuringly though, the relative proportions of
early and late type systems recovered from our classification process
are at least broadly
consistent with those reported by \citet{bui11} in their
analysis of the (lower resolution) GOODS
NICMOS Survey (GNS, \citealt{con11b}; also
\citealt{mor11}).

\section{Statistical Methodology \& Results}\label{method}
Here we begin by introducing our
 stochastic model for the morphological transformation of high
 redshift galaxies, detailing both the tractable
``independent evolution'' case and the intractable
``co-evolution'' case, in Section \ref{stochastic}.  We then proceed to outline the SMC approach
to ABC in Section
\ref{smcabcalg}, and to demonstrate linear regression-based
construction of an optimal summary
statistic--discrepancy parameter pairing for our model in Section
\ref{sumstats}. Finally, in Section \ref{smc} we
compare the performance of SMC ABC against ``tractable likelihood''-based MCMC in the ``independent
evolution'' case and present our ABC-only solution for the more realistic ``co-evolution'' case.

\subsection{Morphological Transformation as a Stochastic
  Process}\label{stochastic}
With the current generation of SAMs yet to offer detailed or reliable predictions for the morphologies of
simulated galaxies \citep{alm07,gon09} we develop here instead a
basic stochastic model for describing the competing processes of
morphological transformation in the early Universe. In this endeavour we are
motivated both by contemporary
observational results and hydrodynamical simulations.  The purpose of
simulation in this study is thus \textit{not} to work forwards through
parameterised approximations for the physical laws of halo accretion, gas cooling,
 and star formation (amongst others) in order to constrain their ``fundamental''
scaling coefficients (as in SAMs), but rather to explore in a rigorous statistical sense the extent to which the
rates of incidence of the key events thought to shape morphological
evolution are jointly constrained by the observed demographics.
Nevertheless, working backwards from these constraints (our posterior probability
densities) one may hopefully achieve insight into the underlying
physical mechanisms, as we discuss in Section \ref{results}.

\begin{figure*}
\hspace{-0.85cm}\includegraphics[width=11.25cm]{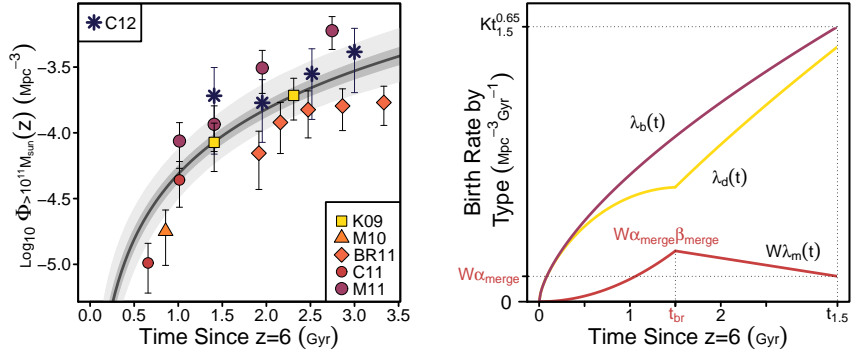}
\caption{(\textit{Left:}) The build up in galaxy number density at
  stellar masses $M_\mathrm{gal} > 10^{11}$$M_\odot$ from
  $z=6$ to $z=1.5$ (in terms of cosmological time since $z=6$)
  synthesised from recent observational determinations of
  $\Phi_{>10^{11}M_\odot}(z)$ from the literature. The K09, M10, C11, and M11
  datapoints shown here are derived from integration of their published Schechter mass
  function fits, while the BR11 datapoints are sourced directly from
  that paper.  The error bar on each indicates the
  1$\sigma$ contribution of cosmic variance for the respective survey and
  bin width (following the recipe of \citealt{mos11}). The dark, medium, and light grey bands plotted underneath
  represent our
  (pointwise) median, 1$\sigma$, and 3$\sigma$
  credible intervals, respectively, on the evolving mean number density.
  These are derived from the joint posterior densities of $K$ and $\gamma$ under our
  non-homogeneous Poisson birth process model, $\lambda_b(t) = 10^K
  t^{\gamma}$; the median curve shown here corresponds roughly to $\gamma=0.65$.
  (\textit{Right:}) Illustration of the birth rate by type in our
  model (cf.\ Section \ref{stochastic}).  The rate at which
  sub-$10^{11}$$M_\odot$ galaxies are promoted above this mass
  threshold by merging is taken as $W$ times our $M_\mathrm{gal}>10^{11}$$M_\odot$
  merger rate, $W\lambda_m(t)$; leaving the rate of promotion (by star formation) of Type
  \textsc{iii} disks, $\lambda_d(t)$, as the remainder with respect the
  total birth rate, $\lambda_b(t)$.   \label{birthprocess}}
\end{figure*}

As a starting
point for our model we suppose that the arrival of galaxies at the top end of
the high redshift stellar mass function may be faithfully represented as a
non-homogeneous Poisson birth process with an underlying rate, $\lambda_b(t)$, increasing
as $10^K t^{\gamma}$ from
a zero baseline
at $z=6$.  Thus, on average (i.e., over an infinite volume), $\Phi_{>10^{11}M_\odot}(z) = \int_0^{t_z}
\lambda_b(t) dt = 10^K \frac{t_z^{\gamma+1}}{\gamma+1}$ (modulo
the impact of merging amongst $M_\mathrm{gal}>10^{11}$$M_\odot$
systems, which is
intrinsically rare at these redshifts, i.e., negligible in this
context; cf.\ \citealt{man11} and our discussion in Section \ref{majormerger}).  In the
lefthand panel of Figure \ref{birthprocess} we illustrate the build up
in number density at $M_\mathrm{gal}>10^{11}$$M_\odot$ over the interval $z\sim1.5$ to 6 synthesised from observations in the
GNS (\citealt{mor11}, M11), the MOIRCS\footnote{MOIRCS: Multi-Object InfraRed Camera and Spectrograph.} Deep Survey
(\citealt{kaj09}, K09), the NEWFIRM\footnote{NEWFIRM: NOAO
  (National Optical Astronomical Observatory) Extremely Wide-Field InfraRed iMager.} Medium-Band
Survey (\citealt{mar10}, M10; \citealt{bra11}, BR11), the UDS
(\citealt{capu11}, C11), and the EGS (the present study,
C12).\footnote{Though redshift may be the more familiar baseline for
  many observational astronomers we have instead adopted here a scale
  of time (since $z=6$, in Gyr) for the horizontal axis of this plot.
This is because time, rather than redshift, forms the natural
evolutionary variable of the stochastic processes described in our
morphological transformation model.  The top right panels of Figures
\ref{mcmcposteriors}, \ref{abcposteriors} and \ref{coevposteriors} are
marked with both, however, as a convenient reference for the
appropriate 
conversion under our assumed cosmology.}
Interestingly, where
their redshift baselines overlap a number of these rival $\Phi_{>10^{11}M_\odot}(z)$
determinations exhibit surprisingly
large discrepancies with regard to their respective cosmic
variance uncertainties (marked as 1$\sigma$
error bars in Figure \ref{birthprocess} following the recipe of
\citealt{mos11}).  As highlighted by \citet{bra11} such discrepancies
may well arise from the systematic errors inherent in
SED-based stellar mass computation (owing to degeneracies between the various template
libraries), and we suspect this to be the case here.

To estimate our birth rate parameters, $K$ and $\gamma$, we thus perform
standard MCMC exploration of the relevant posterior probability
density space under a likelihood model in which the datapoints from
each of the above-listed studies are assumed subject to a common
systematic bias in addition to cosmic variance.  The prior magnitude of this
systematic bias component (in dex) is treated as normally-distributed with
$\mathcal{N}(0,0.1^2)$ for each survey.  Our respective priors on $K$ and $\gamma$ are both
Uniform, with the former non-informative and the latter standard (i.e., bound
between zero and one).  The joint posterior density for
$\{K,\gamma\}$ thus recovered is roughly bivariate
Normal with
$f_{K,\gamma} \sim \mathcal{N}_\mathrm{Trunc.}([-4.1,0.65]^{\prime},[0.06^2,0.1^2;\rho=0.05];0 < \gamma < 1)$.
For reference we plot the corresponding (pointwise) median, 1$\sigma$, and 3$\sigma$
credible intervals for $\Phi_{>10^{11}M_\odot}(z)$ against
the various empirical determinations shown in Figure \ref{birthprocess}.  Due to the relatively small
cosmic volume probed by
the CANDELS/EGS dataset we do not attempt
to further constrain $K$ and $\gamma$ during our ABC analysis;
instead we treat these two variables as nuisance parameters of our
stochastic model and integrate them out at run-time (see Section
\ref{smcabcalg}).

It is important to note at this point that the marginal posterior
density on the systematic bias in our EGS datapoints favours a
(median) of $+$0.11 dex, suggesting that the B11 stellar masses are systematically \textit{over}-estimated by a corresponding $\sim$0.10 dex
(adopting the $z\sim1.5$ mass function slope of \citealt{mor11}).  
Hence, it is perhaps more appropriate to describe our B11 CANDELS/EGS
dataset as an
$M_\mathrm{gal}\gtrsim10^{11}$$M_\odot$ sample, stressing the inherent
(systematic) 
uncertainty in SED-based stellar mass selection (arising primarily from the [uncertain] choice of stellar population synthesis
model/code used to construct the underlying SED template library; cf.\ \citealt{muz09}).

We next suppose that each galaxy
arrives at the top end of the high redshift stellar mass function as
either a (star-forming)
late-type disk (Type \textsc{iii}) or an ongoing major merger
(Type \textsc{iv})---a simplifying assumption which serves to reduce markedly the required dimensionality of our
model, yet which is also consistent with the present state
of knowledge on this topic.  In a recent empirical census
of rest-frame optical morphology amongst the sub-$10^{11}$$M_\odot$
population at $1.5 < z < 3.5$ \citet{cam11} were unable to identify a
single unambiguous spheroid beyond $z \approx 2.2$ in their sample
from the ERS (and see also
\citealt{con11a} for a similar result).  Amongst the
small fraction ($\sim$20\%) of sub-$10^{11}$$M_\odot$ spheroids
discovered in their sample at later epochs only one was found to be
actively star-forming---leaving dry merging as perhaps the only feasible (but also
unlikely, cf.\ \citealt{lin10,cho11}) mechanism for sub-$10^{11}$$M_\odot$
spheroids to thus move above this threshold mass without transitioning
through a standard Type \textsc{iv} phase.  Meanwhile, contemporary hydrodynamical
simulations have demonstrated the theoretical potential for high
redshift disks at $10^{10.5}$-$10^{11}$$M_\odot$ to
sustain immense rates of star formation fueled by
cold flow gas accretion \citep{dek09,bro09,gen10} while avoiding
secular bulge assembly through the wind-driven disruption
of clump instabilities \citep{gen10,hop11}---ensuring their rapid
transition to the high mass regime intact as Type
\textsc{iii} systems.

The probability of birth as a Type \textsc{iv} ongoing merger is
estimated in our model as $W$ times the ratio of the instantaneous merger rate amongst our
$M_\mathrm{gal}>10^{11}$$M_\odot$ population,
$\lambda_m(t_\mathrm{birth})$, to the corresponding instantaneous birth rate, $\lambda_b(t_\mathrm{birth})$,
as shown in the righthand panel of Figure \ref{birthprocess}.
This factor, $W$, represents another nuisance parameter of our model
corresponding to the ratio by number density of galaxies in
 such a mass range that a single major merger could promote them above $10^{11}$$M_\odot$ to
those already beyond this threshold.  (There is an implicit assumption
here that the merger rate does not evolve significantly with mass over
this small baseline.)  According to the shape of the stellar mass
function at these redshifts (e.g.\ \citealt{bra11,mor11}) we estimate
$W \approx 0.5 \pm 0.2$.  Important to note is the fact that the merger
rate so defined, $\lambda_m(t)$, is strictly that of galaxies already at the top end of the stellar mass function---i.e., we are effectively
eliminating the contribution to the observed merger fraction from
galaxies that were sub-$10^{11}$$M_\odot$ prior to their most recent
encounter.  With
$\lambda_b(t)$ the total birth rate and $W\lambda_m(t)$ the birth rate
of Type \textsc{iv} mergers the corresponding birth rate of Type \textsc{iii} disks,
$\lambda_d(t)$, is simply the arithmetic difference,
$\lambda_b(t)-W\lambda_m(t)$ (as indicated in the righthand panel of Figure \ref{birthprocess}).

As for the total birth rate described earlier, $\lambda_b(t)$, merging in our model is characterised as a non-homogeneous
Poisson process, with a unique rate by volume of
\[
\lambda_m(t) =\left\{
\begin{array}{ll}
\frac{\alpha_\mathrm{merge}\beta_\mathrm{merge}}{{t_\mathrm{br}}^2}t^2
& \mbox{for } 0 \le t
\le t_\mathrm{br},\\
\left\{ \begin{array}{l} \alpha_\mathrm{merge}\beta_\mathrm{merge}- \\
  \frac{(t-t_\mathrm{br})\alpha_\mathrm{merge}(\beta_\mathrm{merge}-1)}{t_{1.5}-t_\mathrm{br}} \end{array} \right.
  & \mbox{for } t_\mathrm{br} < t \le
t_{1.5}.
\end{array}
\right.
\]
Here $\alpha_\mathrm{merge}$ represents the baseline merger rate (in
units of Mpc$^{-3}$Gyr$^{-1}$) at our lowest redshift, $z=1.5$, with
$\alpha_\mathrm{merge}\beta_\mathrm{merge}$ the peak cosmological merger
rate for massive galaxies at $t_\mathrm{br}$.  This latter model
parameter, $t_\mathrm{br}$, thus dictates a point of phase transition (or ``break'') beyond which
the merger rate by volume must ultimately decrease (with
\textit{increasing redshift}) back to zero at $z=6$ (the time origin of our
model) at least as fast as the total number density of galaxies
itself, lest the
specific merger rate (per galaxy), $\frac{\lambda_m(t)}{\Lambda_b(t)}$, become
asymptotically infinite.  Here we have chosen for simplicity a fixed,
marginally sufficient decay rate for $\lambda_m(t)$
above this transition redshift of $t^{2>\gamma(\approx0.65)+1}$.  One
possible extension of our model, which may well be worthwhile in the future
if/when a larger demographic dataset for $z\sim1.5$ CANDELS sources
becomes available, would be to treat this pre-$t_\mathrm{br}$ decay
rate as a free parameter of the fit.

Previous empirical studies have argued alternately
that the massive galaxy merger rate is either very near constant
\citep{der11,wil11,lot11,man11} or markedly increasing with redshift
\citep{con03,con09,lop09,der09,blu09,blu11} from
the local volume out to $z\sim1.5$; and even less consensus exists regarding its
behaviour to $z\sim 3$ and beyond \citep{blu11,man11,law11,wil11}. The
intrinsic clumpiness of galaxy-scale star-formation at the observed optical (i.e., rest-frame
UV) wavelengths of many of these studies (most non-WFC3) has proved a persistent source of
uncertainty, introducing substantial ambiguity into the
interpretation of those morphological signatures otherwise indicative
of recent merging locally (cf.\
\citealt{con03,lop09}).  Uncertainties concerning the fraction
of apparent close pairs to ultimately merge \citep{lot08} and
the visibility timescales
of the resulting post-merger tidal features \citep{lot10} have only
compounded these
difficulties.  (Important to note is that the full demographic analysis
performed in the present study permits a simultaneous, non-degenerate constraint of
the latter unknown, which is a significant advantage of this particular
mode of analysis.)

\begin{figure*}
\hspace{-0.85cm}\includegraphics[width=10.54cm]{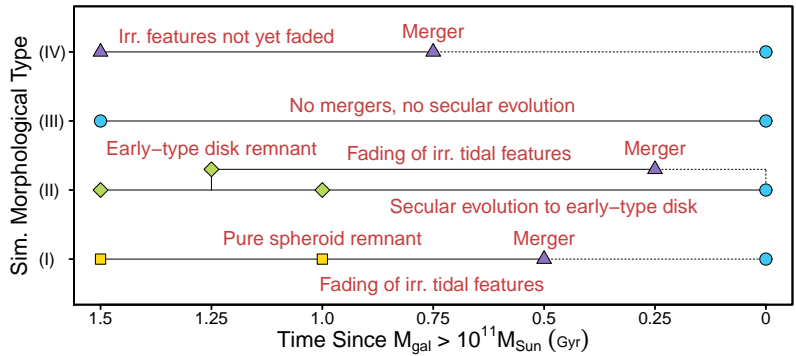}
\caption{Schematic illustration of the five characteristic \textit{pathways of
  high redshift morphological transformation} permitted under the
  stochastic model described in this paper (shown over an arbitrary timeline
  of 1.5 Gyr from ``birth'' to ``observation'' with an assumed Type
  \textsc{iii} birth class).  Solid lines are used to
  mark the fundamental pathways necessary to reach a given
  evolutionary state, whereas dashed lines
  allow for a range of possible degenerate evolutionary histories prior to the
  most recent merger (the details of which are inconsequential to the
  final state achieved).  The secular evolution pathway
  to Type \textsc{ii} status and the null evolution pathway to Type
  \textsc{iii} status are the only branches for which one could not
  substitute Type \textsc{iv} as the birth class here. \label{pathways}}
\end{figure*}

Our
only inflexible constraints on the tuneable
parameters of $\lambda_m(t)$ here are thus that
$\alpha_\mathrm{merge}$ is, of
course, strictly positive, $\beta_\mathrm{merge}$ is greater
than or equal to one, and $t_6(=0) \le t_\mathrm{br} \le t_{1.5}$.  Hence we adopt only weak priors specified as: \textsc{(i)}
 a T distribution in log${}_{10}$-space for $\alpha_\mathrm{merge}$ [with
 $\mu=-4$, $\Sigma=0.5$, and 10 degrees of freedom, truncated to a key
 region of interest at a lower bound of $-5.5$ and an upper bound of
 $-2.5$]; \textsc{(ii)} a Beta distribution in log$_{10}$-space for $2\beta_\mathrm{merge}$
 [with shape coefficients, 1 and 4, favouring a smaller peak-to-baseline
 ratio over a higher one]; and \textsc{(iii)} a Beta distribution for $t_\mathrm{br}$
 (as a fraction of $t_{1.5}$)
 [with shape coefficients, 2 and 1, favouring a break closer to
 $z\sim1.5$ than $z\sim6$].  The
 grey-shaded tiles and histograms in Figures \ref{mcmcposteriors},
 \ref{abcposteriors}, and \ref{coevposteriors} offer graphical
 representations of these prior densities.

Since, as mentioned earlier, dry-merging appears to be remarkably
uncommon in the early Universe---cf.\ the rapidly declining fraction of red-red pairs with
increasing redshift \citep{lin10,cho11,kam11} and the overall paucity of passive galaxies, in general, above $z \sim
2$ \citep{bra11,whi11,wuy11}---we assume that \textit{all}
mergers to occur under our model are gas-rich and therefore generate  a distinctive post-merger morphology with
irregular tidal features \citep{elm07b,lot10}.  We model the (observed $H$-band)
visibility timescale of these features according to
a Gamma distribution with scale,
$1+100\tau_\mathrm{Irr\ morph}$, and shape coefficient, 100, for
$\tau_\mathrm{Irr\ morph}$ in Gyr; thereby allowing an $\sim$0.1 Gyr interquartile spread to account for some intrinsic variation in
the cold gas fraction (and thus the merger-to-stable-remnant
transition time; \citealt{lot10}) across the galaxy population sampled.  Inspired
by contemporary hydrodynamical simulations of gas-rich mergers \citep{lot10} we choose
our prior density on $\tau_\mathrm{Irr\ morph}$ to favour timescales on the order
of $0.2$ to $0.7$ Gyr (though permitting, at much lower prior
density, the possibility of even $>$1 Gyr timescales;
\citealt{con09b}) by adopting a
Beta distribution with shape coefficients, 3 and 5, on this parameter
divided by 1.5.  Upon
fading of these post-merger tidal features we suppose the final
remnant may assume either a Type \textsc{i} (pure spheroid) or Type \textsc{ii}
(spheroid-plus-disk) morphology with the probability of the latter outcome a tuneable parameter, $P_\mathrm{Sph+D\ remnant}$. Given
the ongoing debate within the hydrodynamical modelling community
regarding the relative frequency of mergers conducive to disk reformation at these
epochs (e.g.\ \citealt{rob06,bou11}) we adopt a
Beta distribution prior on $P_\mathrm{Sph+D\ remnant}$ with
shape coefficients, 1 and 3, favouring Type \textsc{ii} production in less
than one in every four mergers.

The second pathway of morphological transformation permitted under our
model is that of secular evolution of Type \textsc{iii} disks to Type
\textsc{ii}, the theoretical mechanism proposed to drive this process
being the inwards migration of massive star-forming
clumps as encountered in certain hydrodymical simulations
\citep{bou07,elm08,dek09}.  We again model this process stochastically
via a Gamma distribution with scale, $1+50\tau_\mathrm{sec\
  ev}$, and shape coefficient, 50, for $\tau_\mathrm{sec\
  ev}$ in Gyr; thereby inducing an intrinsic spread in this variable at run-time intended to mimic the impact of natural
diversity in the structure and kinematics of high redshift disks.  Though inwards migration has been well
publicised as the favoured hypothesis of the
SINS\footnote{SINS: Spectroscopic Imaging survey in the Near-infrared
  with SINFONI [Spectrograph for INtegral Field Observations in the Near Infrared].}
team to explain the characteristic morphologies and SEDs of the
clump population hosted amongst members of their pioneering $z\sim2$ survey \citep{for11,gen11}, as noted
earlier the most recent hydrodynamical
simulations incorporating the effects of wind-driven mass loss, at least in the
$10^{10.5}$-$10^{11}$$M_\odot$ regime, indicate that
 a large fraction of these clumps may be too short-lived to migrate successfully into a
central bulge \citep{gen10,hop11}.  We therefore adopt such a prior on
the timescale for secular bulge formation as to allow a full range of scenarios from
rapid growth on sub-Gyr timescales (implying that many of our
Type \textsc{iii} disks should transition to Type \textsc{ii} before
$z\sim1.5$) to incredibly slow growth on up to 10 Gyr timescales (implying none should).
Mathematically we represent our prior density on this parameter via a
Uniform distribution in log${}_{10}$-space bounded between -1 and 1.

Figure \ref{pathways} illustrates schematically the five distinct
pathways of morphological transformation permitted under the
above-specified model.  We note for reference that Figures \ref{mcmcposteriors},
 \ref{abcposteriors}, and \ref{coevposteriors} offer graphical
 representations of the prior densities on all our model parameters.

\subsubsection{Simulation from our Stochastic Model}\label{simulation}
Having outlined above the principle details of our stochastic model for high
redshift morphological transformation we now describe our
corresponding procedure for simulating from this model under two distinct
paradigms---``independent evolution'' and ``co-evolution''---the
likelihood of the observed
data under a given
set of model parameters being tractable in the former and intractable
in the latter.  [Our derivation of the ``independent evolution''
likelihood function is given in the Appendix to this paper.]

\paragraph{The ``Independent
Evolution'' Case}
As the name suggests in the ``independent evolution'' case we suppose
that neither the birth nor
morphological transformation history of any galaxy are ever coupled to those of another.  Simulation from our
stochastic model under this assumption for a given set of input
parameters is then simply a matter of applying the above probabilistic transition
rules to generate one-by-one a mock morphology at the observed
redshift for each object in our benchmark sample as follows.

First, the birth time of the galaxy at hand (i.e., the epoch at which
its stellar mass finally exceeds $10^{11}$$M_\odot$)
is drawn from the interval $t_{6} (=0) \le t_\mathrm{birth} \le t_\mathrm{obs}$ according to
the waiting time distribution dictated by its assumed
(increasing-rate, non-homogeneous)
Poissonian form (as derived in the Appendix to this paper).  The birth class is then assigned as either Type
\textsc{iii} or Type \textsc{iv}, with the probability of the latter
given by $\frac{W
  \lambda_m(t_\mathrm{birth})}{\lambda_b(t_\mathrm{birth})}$.  To
compute this ratio we must also sample a value for each of the nuisance
parameters, $K$, $\gamma$ and $W$, according to
$f_{K,\gamma} \sim
\mathcal{N}_\mathrm{Trunc.}([-4.1,0.65]^{\prime},[0.06^2,0.1^2;\rho=0.05];0 < \gamma < 1)$ and $f_W \sim \mathcal{N}_\mathrm{Trunc.}(\mu=0.5,\sigma=0.2;W
> 0)$ respectively (where $\mathcal{N}_\mathrm{Trunc.}$ represents the
truncated Normal distribution).  The number of mergers, $n_\mathrm{merge}$, experienced
between birth and observation is then drawn
from the Poisson distribution specified by rate, $\Gamma_m^\ast =
\int_{t_\mathrm{birth}}^{t_\mathrm{obs}}
\frac{\lambda_m(t)}{\Lambda_b(t)}dt$.  If $n_\mathrm{merge} \ne 0$ the corresponding epoch
of last major merger is identified by
sampling from the relevant waiting time
distribution (also derived in the Appendix).  The manifest duration of the resulting post-merger (Type \textsc{iv}) irregular state is
then drawn directly from the Gamma distribution with scale, $1+100\tau_\mathrm{Irr\ morph}$, and shape  coefficient, 100; and if encompassed within the
remaining time until observation the galaxy is assigned either
Type \textsc{i} or Type \textsc{ii} morphology, with the probability of the latter
set by $P_\mathrm{Sph+D\ remnant}$ (otherwise it finishes the simulation
as a Type \textsc{iv}).  Finally, galaxies born as Type \textsc{iii}
disks and experiencing no major
mergers may yet evolve to Type \textsc{ii} via secular evolution, determined likewise by
comparing an evolutionary period drawn from the Gamma distribution
with scale, $1+50\tau_\mathrm{sec\ ev}$, and shape  coefficient, 50, against the
time
available between birth and observation.

Simulation from our model is thus inherently stochastic---i.e., the
internal assignment of birth times, most recent merger times, and so
on (and thereby the
output assignment of final morphologies) will vary from run to run
at fixed input.  In the SMC approach to ABC \citep{cho02,del06,sis07,dro10}
 the effects of this stochasticity are accounted for in an efficient,
 consistent manner through the iterative application of the
key rejection and resampling/refreshment steps described in Section \ref{smcabcalg}.

\begin{figure*}
\hspace{-0.15cm}\includegraphics[width=16cm]{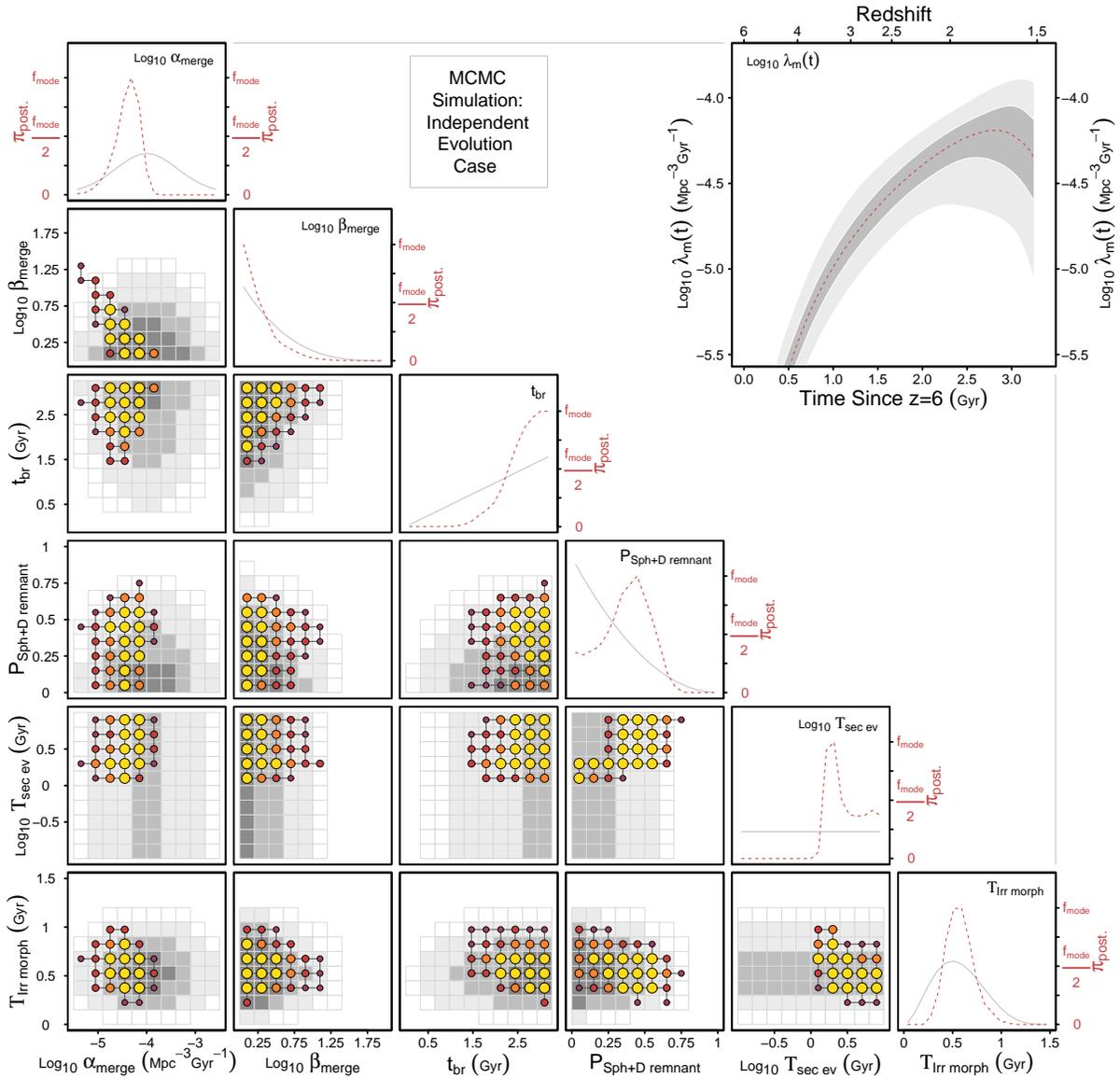}
\caption{Benchmark posterior probability densities for the key
  parameters of our stochastic model of morphological transformation
  at high redshift
  (in the ``independent evolution'' case) recovered from
  ``tractable likelihood''-based MCMC simulation.  In each of the main
  diagonal panels we compare the marginal posterior density of a single
  parameter (in red) against its prior (in grey), while in each of
  the off-diagonal panels below we extend this comparison to the joint
  density formed by pairing that parameter against
  one of its peers.  For the latter visualisation we employ
  a lattice of variable-sized points to
  trace the MCMC posterior on a
  scale of 1, 2.5, 7.5, and 15 times some appropriate baseline probability density, while grey-shaded tiles map the corresponding prior on
  an identical scale.  In the upper right panel we plot
  the
  (pointwise) 1$\sigma$ and 3$\sigma$ credible intervals and median curve (in dark
  grey, light grey, and red respectively) for the $M_\mathrm{gal}>10^{11}$$M_\odot$ merger rate,
  $\lambda_m(t)$, deriving from our (joint, marginal) posterior densities on
  $\alpha_\mathrm{merge}$, $\beta_\mathrm{merge}$, and $t_\mathrm{br}$. \label{mcmcposteriors}}
\end{figure*}

As noted earlier a characteristic feature of our model in the
``independent evolution'' case is that the likelihood, $P(\bm{y}|\bm{\theta})$, of the
observed data under a given set of input parameters is, in fact, tractable (whereas in the ``co-evolution''
case it is not).  For expository purposes this allows us to
reconstruct via standard (``tractable likelihood''-based)
Bayesian computational methods (namely, MCMC) the ``true'' posterior probability density of our
model parameters (modulo the inherent variance of MCMC simulation) as
a benchmark for comparison against our ABC results.  The derivation of
this likelihood function is rather involved so we
present details
separately in the Appendix (along with a description of the MCMC
scheme employed).  The resulting ``true'' posteriors are, however, presented
here in Figure \ref{mcmcposteriors} for reference during our
discussion of summary statistics in Section \ref{sumstats} and the accuracy of
our ABC posteriors in
Section \ref{smc}.

\paragraph{The ``Co-Evolution'' Case}\label{coevsection}  Though it ensures
the likelihood tractability required for our demonstration of the
robustness of SMC ABC with
respect to MCMC in Section \ref{smc} below our initial assumption that
galaxy evolution proceeds independently across our entire
sample may not be physically realistic.  A number of recent studies probing high redshift clusters and proto-clusters
out to
$z\sim1.5$-3
\citep{doh10,hat11,pap11,spi11,tan11} and beyond \citep{cap11,car11}
have presented evidence to suggest that the evolutionary histories of
intermediate mass galaxies within these early
over-densities are in fact highly correlated such that they
consistently achieve peak star formation earlier
than their counterparts of similar mass in the field. Indeed such early biasing by environment of star formation and mass
accretion constitutes a fundamental prediction of hierarchical
formation theory under $\Lambda$CDM
\citep{spr05,ove09} and should already be manifest in the spatial
distribution of the first generation of (re-)ionising sources
\citep{kra06}.  As usual though, empirical results for galaxies at
the top end of the stellar mass function remain limited due to
the intrinsic rarity of these systems.

  The
beauty of ABC, of course, is that it permits the study of
arbitrarily complex stochastic models irrespective of likelihood
tractability, allowing one to relax such simplifying assumptions as that of ``independent
evolution'' in the present example.
In this Section we thus
outline a ``co-evolution'' case of our model in which a physically
plausible coupling is introduced into the formation
times of galaxies in close pairs and small groups, and in Section \ref{smc} we explore
the impact of this coupling on our ABC posteriors.

\begin{figure*}
\hspace{-0.85cm}\includegraphics[width=10.54cm]{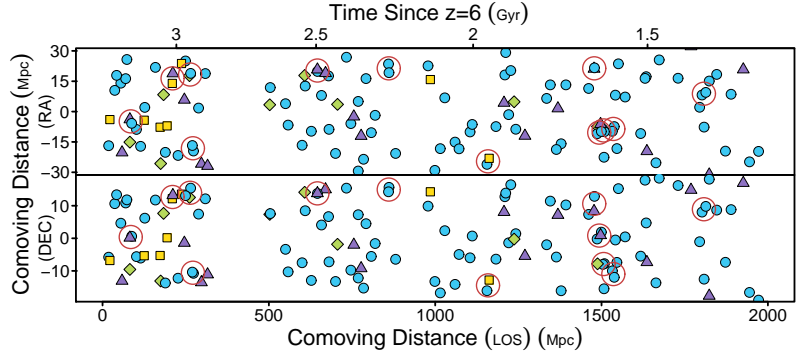}
\caption{The spatial distribution of massive
  ($M_\mathrm{gal} > 10^{11}$$M_\odot$) galaxies in our B11 (CANDELS/EGS)
  sample at $1.5 < z < 3$, projected in terms of comoving distance onto the LOS-RA and LOS-DEC
  planes.  The red circles overplotted highlight our eleven pairs
  and one threesome of galaxies neighbouring within 2.5 Mpc.  The
  colour/symbol code for galaxy types employed here is
  identical to that of Figure \ref{morphologies}.  For reference we
  also mark the LOS axis on a scale of time since $z=6$ (the origin
  point of our model).  [See the text for a comment on
  the apparent void at $\sim$400 Mpc.]\label{radec}}
\end{figure*}

In Figure \ref{radec} we illustrate the nature of spatial clustering
amongst the massive ($M_\mathrm{gal} > 10^{11}$$M_\odot$) galaxies of
our B11 (CANDELS/EGS) sample at $1.5 < z < 3$, representing
their 3D distribution in observed right ascension (RA),
declination (DEC), and redshift via 2D projections in comoving distance along
the line of sight (LOS) and the axes of RA and DEC,
alternately.  Neighbouring systems separated by no more than 2.5
Mpc---a conservative linking scale for proto-group-sized over-densities
in the early Universe (cf.\ \citealt{cap11} and references
therein)---are marked accordingly and highlight the diversity of high redshift
``environments'' in the survey volume, with eleven simple
pairings and one threesome identified at the adopted linking scale and
a majority of relatively isolated systems.  Perhaps the most striking feature on first
  inspection of this plot, however, is the apparent void at $\sim$400
  Mpc distance from the lower bound of our sample at $z\sim1.5$.
  Examination of the B11 photometric redshifts for the full EGS, however,
   reveals no indication of this under-density extending across the
   wider field, reassuring us that it is most likely an imprint
  of cosmic variance (cf.\ \citealt{tre08,dri10}) within the small
  volume probed by the present
  dataset---and not the result of a
  systematic bias in the adopted SED fitting
  algorithm, for instance.\footnote{Analysis of the (much larger) COSMOS
    photometric redshift catalog \citep{ilb10} confirms that
    under-densities of this magnitude indeed arise frequently amongst
    the massive/most-luminous galaxy population at these redshifts.  In particular,
    for the optically-luminous (i.e., rest-frame $Z < -23.6$ mag)
    members 
    of that catalog at $1.5
    < z < 3$, which appear at comparable number
    density to the $M_\mathrm{gal} > 10^{11}$$M_\odot$ population from the
    EGS, a maximum LOS separation of 150 Mpc or more 
  occurs (roughly) once for every three random placements of the CANDELS/EGS
  observational footprint within the COSMOS field.}

To explore the impact of small scale clustering on the ``birth''
time distribution as defined in our stochastic model for morphological
transformation we refer to the publicly-available
mocks from the
\citet{del07} SAM embedded in a small volume of the Millennium
Simulation\footnote{URL[http://galaxy-catalogue.dur.ac.uk:8080/Millennium/].}
of comparable size to that probed by the B11 dataset.\footnote{Recall that for the purposes of the present
  analysis we treat the time of ``birth'' in
  our model as the epoch at which a galaxy first reaches the top end of the high redshift stellar mass function, whether
via star formation or merging.  To identify this point in the
\citet{del07} mocks one must follow back the linked progenitor tree
accordingly via \texttt{SQL} query.}  In line with our suspicion from Section
\ref{stochastic}  of an $\sim$0.1 dex
bias in the B11 stellar masses our first discovery here is that the number density
of simulated galaxies in these mocks at a cut-off of $M_\mathrm{gal}>10^{11}$$M_\odot$ is far
below that of our CANDELS/EGS sample, but may be brought into reasonable agreement if we revise
our selection down to (at least) $M_\mathrm{gal}>10^{10.9}$$M_\odot$.  Following
this adjustment we identify 12 pairs, 2 threesomes, 3 foursomes, and
even a five member configuration in this $\sim$$7\times10^5$ Mpc$^3$
``snapshot''  of the \citet{del07} SAM at $z\sim1.5$.  Interestingly, whilst we do find a strong
\textit{correlation} between the birth times of galaxies in
close associations---with a median absolute difference of only $\sim$0.3 Gyr
for neighbours within 2.5 Mpc compared against $\sim$0.65 Gyr for
randomly assigned pairs---despite some theoretical expectation there exists negligible evidence in these mocks for a
systematic \textit{bias} in this specific aspect of galaxy formation.  Hence, we do not attempt to
induce one arbitrarily into our stochastic simulations; instead we focus
here simply
on reproducing the observed correlation using the following modified sampling scheme.

Rather than drawing independent birth
times one-by-one for each galaxy in our sample as in the ``independent evolution''
case described above, in the ``co-evolution'' case of our model we
instead generate a complete set of
birth times at the start of the simulation and distribute these
 thereafter with an
environmental dependence.  We
achieve this via the admittedly somewhat \textit{ad hoc} scheme
described below, which we have specifically tailored (through ``trail
and error'' experimentation) to render median absolute birth time differences for both
neighbouring galaxies and random pairings of similar magnitude to
those of the \citet{del07} model.  That is, we do not propose that our
scheme follows in meaningful way the unknown sequence of random physical processes by
which nearby neighbours come to experience similar evolutionary
histories; we simply assert that it mimics faithfully the
imprint of these processes on the coupling of galaxy birth times
``observed'' in this particular reference SAM.

First we draw from the standard Uniform
distribution a primary set of 126 points, one for each galaxy in the B11 CANDELS/EGS dataset,
plus a secondary set of 12 points, one for each close pair or threesome earlier identified (see Figure
\ref{radec})---the latter serving as ``latent variables'' for the coupling
of birth times in these associations.  To each galaxy in a close pair
or threesome we then allocate a single point from the primary
set with selection probability proportional to the inverse square
of distance between that point and the corresponding
latent variable from the secondary set.  The remaining points in the
primary set are then randomly allocated with equal selection
probability to the many isolated
galaxies of our dataset.  Each
point on the interval [0,1] thus assigned is then transformed to a birth time for its matching galaxy
through multiplication by the relevant $\Lambda_b(t_\mathrm{obs})$
(see, for reference, the Appendix to this paper) followed by
inversion to recover $t_\mathrm{birth}$.  Note that by allocating from an initial uniform sample in this manner
we naturally preserve the mean build up rate of $\Phi_{>10^{11}M_\odot}(z)$
specified by our fit of $K$ and $\gamma$ in $\lambda_b(t)$ against the
available observational data from K09, M10, BR11, C11, M11, and C12
(see the lefthand panel of Figure
\ref{birthprocess}).


\subsection{SMC ABC: Sequential Monte Carlo Approximate Bayesian Computation}\label{smcabcalg}
As mentioned in the Introduction, Approximate Bayesian
Computation (cf.\ \citealt{tav97,pri99,bea02,sis07,wil08,csi10,dro10,dro11})
offers a rigorous statistical framework for estimating the posterior probability
densities of key scientific parameters under complex models for which
the likelihood of the observed data may be entirely intractable (thus
prohibiting application of the standard MCMC approach, for example).  To conduct  ABC one
requires only a stochastic model from which the observed data, $\mathbf{y}$, are
believed to be a random draw given some unknown set of intrinsic (true)
input parameters, and a discrepancy measure for the comparison of simulated
data, $\mathbf{y}_s$, against observed,
$\rho(S(\mathbf{y}),S(\mathbf{y}_s))$, typically based on a set of low-order summary statistics, $S(\cdot)$.  Following \citet{dro10} the aim of ABC may be
 stated formally as the recovery of unbiased samples from the
 distribution described by the approximate joint posterior
density,
\begin{equation}\label{abc}
\begin{array}{ll}f(\bm{\theta},\mathbf{y}_s | \rho(S(\mathbf{y}),S(\mathbf{y}_s)) \le
\epsilon_T) \propto &
 \begin{array}{l} f(\mathbf{y}_s | \bm{\theta})
   \pi(\bm{\theta}) \\
\times \mathbf{1}_{\rho(S(\mathbf{y}),S(\mathbf{y}_s)) \le
\epsilon_T} \mathrm{,} \end{array}\end{array}
\end{equation}
where $\bm{\theta}$ represents a vector of unknown model
parameters, $\pi(\bm{\theta})$ the prior density on those parameters,
and $\epsilon_T$ some target tolerance for the specified discrepancy measure between simulated and observed data.  The indicator function $\mathbf{1}_{\rho(S(\mathbf{y}),S(\mathbf{y}_s)) \le
\epsilon_T}$ assumes value unity for simulated--observed dataset pairs
with metric distance below this tolerance, and zero otherwise.

The archetypal scheme for random sampling from the distribution defined
by Equation \ref{abc} is that of rejection ABC (cf.\
\citealt{pri99}) in which one draws a large sample of $N$ trial
parameter vectors, $\bm{\theta}_i$ ($i=1,\ldots,N$), from the prior,
$\pi(\bm{\theta})$, simulates a corresponding
dataset for each, $\bm{y}_s$, and then rejects all $\bm{\theta}_i$ for which the discrepancy between simulated and observed
data exceeds some target tolerance, i.e., $\rho(S(\mathbf{y}),S(\mathbf{y}_s)) >
\epsilon_T$.  That is, \textit{one adopts as an approximation to the
  posterior the complementary set of input parameter vectors (drawn from the prior)
  for which the corresponding simulated (or mock) dataset appears
  ``close'' to the observed}.  In principle,
those regions of parameter space with  greatest probability of
having generated the observed dataset should be the most frequently
represented amongst this approximate posterior sample---modulo two
(possibly large) sources of error.  Namely, (\textsc{i}) Monte Carlo error due
to the limited number of simulated datasets it will be feasible to
generate, and the even-more-limited number of these that will likely
appear ``close'' to the observed dataset; and (\textsc{ii}) the inherent
 error of the likelihood approximation in ABC arising from the gap between 
 ``close'' (as judged by the summary statistic-based discrepancy
 distance) and
 ``equal to'', which cannot (in general)
be made arbitrarily small if at least some simulated
datasets are to be deemed acceptable.  Except in (unrealistically) fortuitous
(or trivial) circumstances in which the prior
is already very close to the posterior, when exploring high
dimensional parameter spaces ($N_\mathrm{par}\gtrsim3$) under
rejection ABC these intertwined sources of error may well force one into an undesirable trade-off
between an impractically low acceptance rate or
an uncomfortably large tolerance.  Thus, much recent
work in the ABC field has been concerned with the development of more
\textit{efficient} alternatives to rejection ABC, involving
sophisticated algorithms to focus the 
sampling of input parameters for the (computationally expensive) data
simulation phase towards regions of increasingly higher acceptance
probability, though in such a manner (or with the appropriate book-keeping) as to avoid biasing the output posterior approximation.

Perhaps the most promising of these is the Sequential Monte Carlo (cf.\
\citealt{liu01,cho02}) approach to ABC \citep{del06,sis07,dro10} which
proposes  to simulate from Equation \ref{abc} step-wise by
evolving a dynamic population of
``particles'' (with each particle representing a single vector of input model parameters) through a
sequence of intermediate distributions characterised by $f(\mathbf{y}_s | \bm{\theta}) \pi(\bm{\theta})\mathbf{1}_{\rho(S(\mathbf{y}),S(\mathbf{y}_s)) \le
\epsilon_t}$ for $t=1,\ldots,T$, indexing a series of non-increasing targets, $\epsilon_t$.  The
two key stages of the SMC algorithm are thus: \textsc{(i)}
rejection of the most discrepant particles under each target; followed
by \textsc{(ii)}
resampling from amongst the least discrepant particles with some
refreshment mechanism applied to maintain particle diversity.  SMC may
therefore also be referred to as ``particle
filtering'' or ``population Monte Carlo'' (PMC; for examples of the recent
adoption of PMC [SMC] in a cosmological context see
\citealt{wra09}, \citealt{kil10}, and references therein).

In this study we
employ for
rejection the self-generating target strategy of \citet{dro10} in which the sequence of incremental targets,
$\epsilon_t$, is chosen on run-time such that a fixed fraction,
$\alpha$, of all particles are dropped at each iteration (herein
$\alpha=0.75$).  We then restore the
particle population to its full operating size, $N$, by resampling
with replacement from amongst the remaining
$(1-\alpha)N$ particles.  Population refreshment is achieved by application of
an MCMC kernel to these replicates.  As in
\citet{dro10} we favour the use of a self-refining, Metropolis-Hastings proposal distribution
based on the current particle sample mean vector, $\bar{\bm{\mu}}$, and covariance
matrix, $\bar{\bm{\Sigma}}$.  For the specific model at hand we adopt a
truncated, multivariate T distribution of degree 10 with the
truncation bounds set by the support of our prior densities (as detailed in Section \ref{stochastic}).  At each iteration of the SMC algorithm this MCMC kernel is run a fixed number of
times, $R$, to (hopefully) produce genuine
refreshment in a fraction, $c$, of the resampled particles
(with some particles, of course, likely to be moved multiple times).
The requisite $R$ is here estimated according to the empirical efficiency, $p_\mathrm{acc}$, of
the previous MCMC kernel as $R =
\frac{\log(1-c)}{\log(1-p_\mathrm{acc})}$.  Note that the
``likelihood ratio'' in the corresponding MCMC acceptance computation is
simply $\mathbf{1}_{\rho(S(\mathbf{y}),S(\mathbf{y}_s(\bm{\theta}_\mathrm{proposed}))) \le
\epsilon_t}$, i.e., all trial particles for which the
simulation produces a mock dataset with discrepancy distance falling below the current target
are assured a non-zero probability of acceptance.\footnote{When using a non-symmetric proposal
distribution as in the present case the ratio of sampling densities
joins, of course, the likelihood ratio and the prior density ratio in computing the full
MCMC acceptance probability.}  Our final target, $\epsilon_T$, is
defined pragmatically (with respect to the limitations of our
computational resources) as that $\epsilon_t$ for which a further SMC
rejection--resampling step would incur more than
$R_\mathrm{max} \sim 100$ applications of this MCMC kernel.

\textit{\textbf{Treatment of Nuisance Parameters}}\ \ A particular
feature of the model adopted in this study is the appearance of
three nuisance parameters---essential inputs of little interest to our
final science goal though known only to a
limited accuracy from previous
studies.  Namely, $\bm{\Theta} = \{ K, \gamma, W \}$
where $f_{K,\gamma} \sim
\mathcal{N}_\mathrm{Trunc.}([-4.1,0.65]^{\prime},[0.06^2,0.1^2;\rho=0.05];0 < \gamma <
1)$  and $f_W \sim \mathcal{N}_\mathrm{Trunc.}(\mu=0.5,\sigma=0.2;W
> 0)$.  Upon each simulation from our model for a single particle,
$\bm{\theta}_i$, we draw a random value for each of $K$, $\gamma$, and $W$ from their respective distributions for use in
that one instance.  Heuristically this amounts to approximating the
integral, $\int_{\Omega(\bm{\Theta})} f(\bm{y}_s|\bm{\theta}_i,\bm{\Theta})f(\bm{\Theta})d\bm{\Theta}$, by a single Monte
Carlo sample, yet produces through the power of the SMC ABC algorithm
(i.e., via inference
over a particle
population \textit{en masse})  a
reasonable approximation (converging asymptotically) to sampling from Equation \ref{abc}.

\subsection{Refinement of Summary Statistics}\label{sumstats}
As explored in a number of recent papers careful selection of
the summary statistic(s) used to evaluate the discrepancy between simulated and observed
data in ABC is of paramount importance to achieving accuracy and
efficiency with this algorithm, whether employed for the purpose of parameter estimation (e.g.\
\citealt{joy08,nun10,fea12}) or, more
challengingly\footnote{Application of ABC to
the problem of Bayesian model choice (cf.\ \citealt{gre09,ton10}) is far from
straightforward as an unfortunate summary statistic selection can lead to disasterously incorrect Bayes factors,
even asymptotically \citep{rob11}.  Recently though, \citet{mar11}
have made substantial progress in this field by establishing necessary
and sufficient conditions on the validity of candidate summary
statistics for this purpose.}, Bayesian model choice
\citep{rob11,mar11,bar11}.  Indeed the same is true for all
approaches to inference from complex
models for which sufficient statistics are unavailable, including those
of ``approximate likelihood'' \citep{woo10} and model emulation
(cf.\ \citealt{bow10} and references therein), and should thus not be thought
of as a unique concern for ABC-based analyses.

A summary statistic may be defined as any mathematic representation of
the original dataset that reduces its effective dimension.  For
example, a single column mean, a list of multiple column means, or even a list
of multiple column means, variances, and higher-order 
moments; though it will rarely be profitable
to carry such a high dimensional statistic as the latter into a full ABC analysis owing to
the direct relationship between Monte Carlo error and summary
dimension (cf.\ \citealt{bea02,fea12}).  For some stochastic
models (including that of the present case study in morphological transformation; cf.\ Section
\ref{stochastic}) the nature of the output data may well be such that
only a few basic modes of summary are of any likely value, while for others
(e.g.\ the Ricker map case study of \citealt{woo10}; and most
SAM-based studies of galaxy formation\footnote{With the choice of
  summary statistic typically (re-)cast in terms of the choice of reference
  dataset in these astronomical studies---namely, whether to
  constrain, for instance, against the luminosity function \citep{bow10,lu11,cir10}, the Tully-Fisher relation \citep{van00,ton11}, the mass-metallicity relation \citep{pip09}, and/or the black hole--bulge mass
relation \citep{hen09}.\label{samstat}}) there may in fact be many in a rich
hierarchy of complexities.  Given that ABC is specifically designed
for use with complex stochastic models with intractable likelihoods it
cannot be expected that the inferential value of any of these
candidate summary statistics will be known \textit{a priori} to the
analyst who must ultimately choose between them---though some may well
be strongly biased or uninformative.  Hence in this Section we
complete our exposition of the ABC technique with a demonstration of
two contemporary approaches to this particular selection problem:
first, we apply Nunes \& Balding's
(2010) two-stage procedure of distributional entropy and MRSSE\footnote{MRSSE:
  Mean square Root Sum of Standard Errors.} minimisation to identify
the optimal choice from amongst a
candidate set of four ``na\"ive''
 summary statistics for our morphological dataset; and second, we
 employ the so-called ``semi-automatic''
  scheme of \citet{fea12} to build an alternative
  set of summary statistics optimised with respect to the recovery of
  posterior means, validating their performance in comparison against the
  former.\\

\textit{\textbf{Minimum Entropy/MRSSE-Based Selection}}\ \ The most natural mode of summary for population demographic
data in the context of extra-galactic astronomy is, of course, by way
of type counts or type fractions in
similar-sized bins of redshift (see \citealt{oes10} and \citealt{bui11} for
recent examples).  Important considerations when compiling such summary
data for the purpose of ABC analysis are then the number and placement
of bins to use and
the weights one should assign to the type counts/fractions observed therein.  As a ``na\"ive'' first attempt at constraining the parameter
space of our model in the ``independent evolution'' case we thus trial four
alternative summary statistics based on progressively finer
subdivisions of the B11 CANDELS/EGS dataset by redshift.
Mathematically, we define this class of ``na\"ive'' summary statistics,
$\bm{S}$, via the generic column vector
\[
\bm{S} \equiv S(\bm{y})  = \{f_{\textsc{i}}^{(1)},f_{\textsc{ii}}^{(1)},f_{\textsc{iii}}^{(1)},f_{\textsc{iv}}^{(1)},\ldots,f_{\textsc{i}}^{(k)},f_{\textsc{ii}}^{(k)},f_{\textsc{iii}}^{(k)},f_{\textsc{iv}}^{(k)}\}^\prime,
\]
with $f_{\textsc{i}}^{(1)}$ the fraction of Type
\textsc{i} galaxies in the first of $k$ redshift bins subdividing
equally the interval $1.5 < z < 3$, $f_{\textsc{ii}}^{(1)}$ the
fraction of Type \textsc{ii} galaxies in the aforementioned bin, and so on.   Adopting equal significance
weights across all bins, we thus establish a complete
discrepancy distance,
\begin{equation}
\rho(S(\mathbf{y}),S(\mathbf{y}_s)) =
\sqrt{(\bm{S}_\mathrm{obs}-\bm{S}_\mathrm{sim})^\prime
(\bm{S}_\mathrm{obs}-\bm{S}_\mathrm{sim})}\mathrm{.}
\end{equation}

\begin{figure*}
\hspace{-0.85cm}\includegraphics[width=11.25cm]{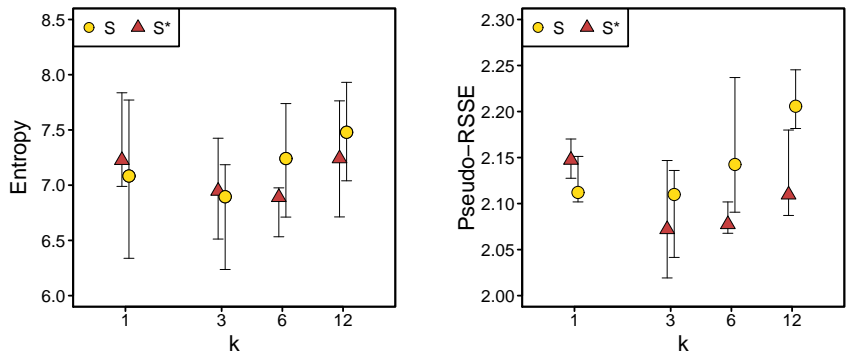}
\caption{Evaluation of candidate summary statistics for model--data comparison across a range of binning schemes
  ($k=1, 3, 6, 12$) for both our ``na\"ive'' $\bm{S}$ and optimised
  $\bm{S}^\ast$ (the latter explained later in this Section) via the twin diagnostics of (\textit{Left:})
  distributional entropy and (\textit{Right:})
   (M)RSSE (cf.\ \citealt{nun10}).  Recall here that a lower posterior
   entropy typically indicates a higher posterior information content,
   and a lower (M)RSSE score a more accurate recovery of the posterior
   mean.  In each instance the marked datapoint reveals the
  median, and the error bars a corresponding 95\% confidence
  interval, recovered from six rounds of rejection ABC
  (i.e., selection of the 100 least discrepant particles out of an initial
  5,000 drawn from the prior density).  Note that as the posterior
  means of our model parameters under the ``independent evolution''
  case studied here have already been well approximated via our earlier (``tractable
  likelihood''-based) MCMC simulation we have employed these directly
  to estimate a pseudo-RSSE, rather than forming an MRSSE
  from repeated ABC runs against simulated datasets ``close'' to the real one as
  in the canonical \citet{nun10} procedure.\label{entropy}}
\end{figure*}

Following the two-stage procedure of \citet{nun10} we begin the evaluation
of our four ``na\"ive'' candidates, $\bm{S}: k= \{1, 3, 6, 12\}$, by computing the fourth-nearest neighbour entropy\footnote{One may note an
  intriguing similarity between the use of fourth (or fifth) nearest
  neighbour-based estimators in both statistical studies of
  distributional entropy and in astronomical studies of large-scale
  environment (cf.\ \citealt{bal06})---though it is unlikely
  there exists an underlying significance to this beyond the desirable error properties
  of the $n\sim4$-5 choice
  \citep{sin03}.} of the
posterior distribution resulting from simple rejection ABC under each---the
goal here being to exploit the (approximate) inverse relationship between entropy and
information in order to identify the most
``informative'' summary statistic with regard to inference of the model parameters
at hand.  In the present round of rejection ABC experiments we
accept only the 100
least discrepant particles of an initial sample of 5,000 drawn from the
prior, and we compute the associated entropy statistic
for each according to the formula,
\[
\hat{H} = \log\left[ \frac{\pi^{N_\mathrm{par}/2}}{\Gamma(N_\mathrm{par}/2+1)} \right] - \psi(4) + \log
  n + \frac{N_\mathrm{par}}{n}\sum_{i=1}^{n}\log R_{i,4}
\]
(\citealt{sin03}; with $N_\mathrm{par}[=6]$ representing the dimension of our model
parameter space, $n[=100]$ the number of accepted particles,
$\Gamma(\cdot)$ and $\psi(\cdot)$ the ``gamma'' and
``digamma'' functions, respectively, and $R_{i,4}$ the fourth nearest
neighbour distance).  There exists, of course, a
certain subjectivity in the choice of scaling for each parameter in
the computation of $R_{i,4}$; one option would be to first standardise all
parameters on the interval $[0,1]$, however, in this case we prefer
instead to standardise against the diagonal matrix of our prior variances, $\bm{V}$,
such that $||\bm{\theta}_i,\bm{\theta}_j||=\sqrt{\bm{\theta}_{i}^\prime
\bm{V}^{-1} \bm{\theta}_j}$.  By repeating the rejection ABC
process six times for each $k$ one may estimate both the
median entropy and matching 95\% confidence interval (from the range)
under that particular binning scheme.  The results of
this analysis are presented in the lefthand panel of Figure \ref{entropy}.

Interestingly,
although one might, at face value, expect a monotonic relationship of
decreasing posterior distributional
entropy with increasing $k$ on the basis that finer binning should break any
false degeneracies in the posteriors recovered from ABC runs with fewer
bins (i.e., in some sense increase the information return), this is
not necessarily true in practice owing to the simultaneous increase in
Monte Carlo error, as the present example demonstrates.
Although we do observe a
slight decrease in distributional entropy upon moving from one to
three redshift bins the opposite is true for our six and
twelve bin trials under this particular
mode of summary (see the lefthand panel of Figure \ref{entropy}).

Following identification of the entropy-minimising $\bm{S}$ (here $k=3$) \citet{nun10} recommend a second
(rather computationally expensive) round of evaluation against the formal
optimality criterion of Mean square Root Sum of Squared
Errors (MRSSE) to ensure that the summary statistic favoured by the
minimum entropy analysis is not likely biased with respect to
recovery of the posterior mean.  In the full \citet{nun10} scheme the
MRSSE score is to be estimated via a new series of
rejection ABC analyses against simulated datasets constructed under
parameter vectors revealed by the original ABC runs for the
entropy-minimising $\bm{S}$ as (likely to be) ``close'' to those responsible
for
the observed dataset.  Since the (marginal) posterior
  mean of each model parameter under the ``independent evolution''
  case studied here has already been well approximated via our earlier (``tractable
  likelihood''-based) MCMC simulation (see Section \ref{stochastic})
  we may take a shortcut to the truth here (and vastly reduce our
  computational burden) by employing these
   directly to estimate alternative pseudo-RSSE scores for
   each of our previous rejection ABC runs.  The results of
  this analysis are presented in the righthand panel of Figure
  \ref{entropy}.  The relationship between
  pseudo-RSSE score and binning $k$
  observed here
  mirrors closely that exposed by our original entropy
  evaluation, validating $\bm{S}:k=3$ as
  the
  optimal choice from amongst our candidate set of ``na\"ive'' summary statistics.

\begin{figure*}
\hspace{-0.85cm}\includegraphics[width=11.25cm]{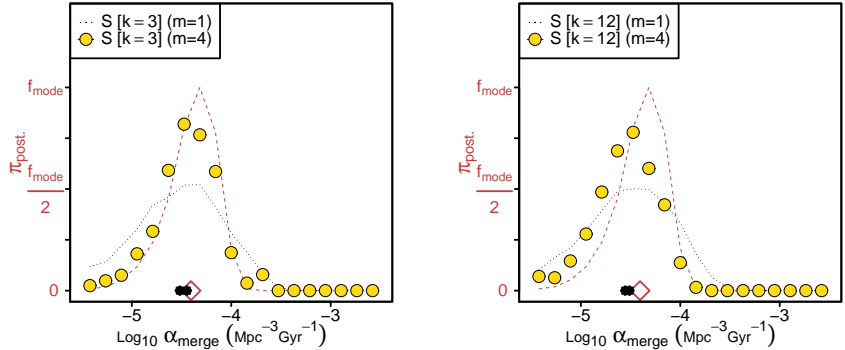}
\caption{Comparison of the marginal posterior density for
  $\alpha_\mathrm{merge}$ in the ``independent evolution'' case of our
  model recovered from
  SMC ABC under two of our ``na\"ive'' summary statistics, $\bm{S}: k=3$ and $\bm{S}: k=12$).  In each case the dotted line
  represents the posterior after one rejection--resampling iteration ($m=1$) of the SMC ABC
  algorithm, the solid datapoints the same after four
  iterations ($m=4$), and the dashed red line the benchmark
  posterior from ``tractable likelihood''-based MCMC.  At the
  bottom of each panel we indicate also the position of the posterior mean
  for this parameter under each of our SMC ABC runs as well as from
  our MCMC
  benchmark (the latter highlighted with a red diamond).\label{summstatsy}}
\end{figure*}

The value of the
above optimisation procedure may easily be appreciated from inspection
of Figure \ref{summstatsy} in which we compare the marginal posterior
for one of our key model parameters,
$\alpha_\mathrm{merge}$,
recovered from full SMC ABC analysis  under $\bm{S}$ with $k=3$
against that for $k=12$.  As in all SMC ABC runs reported herein we use
a population of 10,000 particles, iterated through an adaptive
threshold defined by a rejection rate of $\alpha=0.75$, followed by
MCMC kernel-based resampling with a goal refreshment rate of no less
than $c = 0.90$.  A total of four iterations were achieved under
this scheme
before the limits of our computational resources were reached (at $R > R_\mathrm{max}[=100]$ required applications of the MCMC
kernel for such a level of refreshment).  Although both (marginal) posteriors
appear rather similar after only one iteration it is evident by the fourth
(and final) iteration that the $\bm{S}:k=3$ statistic has produced a particle
population tracing far more faithfully the density of our benchmark
(``tractable likelihood''-based) MCMC simulation than its
$\bm{S}:k=12$ counterpart.

\textit{\textbf{The ``Semi-Automatic'' Scheme}}\ \  In an important contribution to the ABC literature
\citet{fea12} have recently demonstrated that the optimal summary statistic for estimation
of model parameters under quadratic loss (i.e., optimality with
respect to the recovery of posterior means) is simply the
conditional expectation function, $E(\bm{\theta}|\bm{y})$.  As a
direct consequence the authors were thereby able to propose and justify a
regression-based algorithm for the direct construction of
well-behaved summary statistics, allowing one (in principle)
to by-pass
the above process of searching through a ``na\"ive'' set of
often unsatisfactory candidates.
 To implement their so-called ``semi-automatic'' scheme one must first generate a ``reasonably large'' sample of model
 parameter--simulated dataset pairs spanning a ``relevant volume'' of
 parameter space.  In lower dimensional analyses one
 may simply draw this sample directly from the prior, though the posterior density from a trial ABC run with some ``na\"ive''
 summary statistic will generally serve as a superior starting point.
 Least-squares-based fitting to this reference dataset of the relation, $\theta_i = \beta_0^{(i)}
 + \bm{\beta}_1^{(i)}f(\bm{y}) + e_i$, for each model parameter,
 $\theta_i \in \bm{\theta}$, yields the optimal summary statistic,
 $\bm{S}^\ast =  \bm{\beta}_0 + \bm{\beta}_1 f(\bm{y})$.\footnote{In
   fact, as noted by \citet{fea12}, since in ABC analysis we are only interested in
   the difference,
   $\bm{S}^\ast_\mathrm{obs}-\bm{S}^\ast_\mathrm{sim}$, the vector,
   $\bm{\beta}_0$, may well be omitted from this above definition.}  Here $e_i$ denotes a symmetric
 error term of zero mean and $f(\bm{y})$ some
 vector-valued function of the data, which for
 $\bm{y} \in \mathds{R}^p$ will be typically of the
 form $f(\bm{y})=\bm{y}$,
 $f(\bm{y})=\{\bm{y},\bm{y}^2\}^\prime$, or similar---the lack of a
 universally appropriate algorithm for defining this
 regression function being the chief cause for classification of the above scheme
 as ``semi-'' rather ``fully'' automatic.

Since the raw output, $\bm{y}$, from our stochastic model for high redshift morphological
transformation is, in fact,
  multinomial (rather than real-valued) we adopt
  here (for the purposes of computational efficiency) a modified regression function of form, $f(\bm{y})=S(\bm{y})$, with $S(\cdot)$
 denoting as above the compilation of type fractions in fixed bins of
 redshift.  Under this adaptation of the \citet{fea12} approach the
 magnitude of each component in each fitted $\bm{\beta}_1^{(i)}$ may be
 considered a
  weight for the importance of that type fraction
 and redshift bin in estimating the corresponding ($i$-th) model parameter.  As in our earlier computation of fourth-nearest neighbour
 distances we employ our prior
 variance matrix, $\bm{V}$, to establish the full discrepancy measure under this new
 summary statistic\footnote{Another (well-motivated) alternative choice of scaling here
   would be the sample covariance matrix of the posterior particle
   population from our earlier run of SMC ABC
 under $\bm{S}:k=3$.},
\[
\rho(S^\ast(\mathbf{y}),S^\ast(\mathbf{y}_s)) =
(\bm{S}^\ast_\mathrm{obs}-\bm{S}^\ast_\mathrm{sim})^\prime
\bm{V}^{-1}(\bm{S}^\ast_\mathrm{obs}-\bm{S}^\ast_\mathrm{sim})\mathrm{.}\]
Fitting of $\bm{\beta}_0$ and $\bm{\beta}_1$ was
achieved here by application of the
\texttt{glm} and \texttt{step} routines in \texttt{R} to the 100 least discrepant particles from each of the six
rejection ABC runs conducted earlier under our ``na\"ive'' summary
statistic for $k=3$ (resulting in a full calibration sample of
600 model
 parameter--simulated dataset pairs).  Notably, the \texttt{step} routine in \texttt{R} makes use of the AIC (Akaike Information Criterion) statistic to
 restrict each fit to only those elements of $\bm{S}$ contributing
 significantly to the prediction of $\theta_i$.

\begin{figure*}
\hspace{-0.85cm}\includegraphics[width=11.25cm]{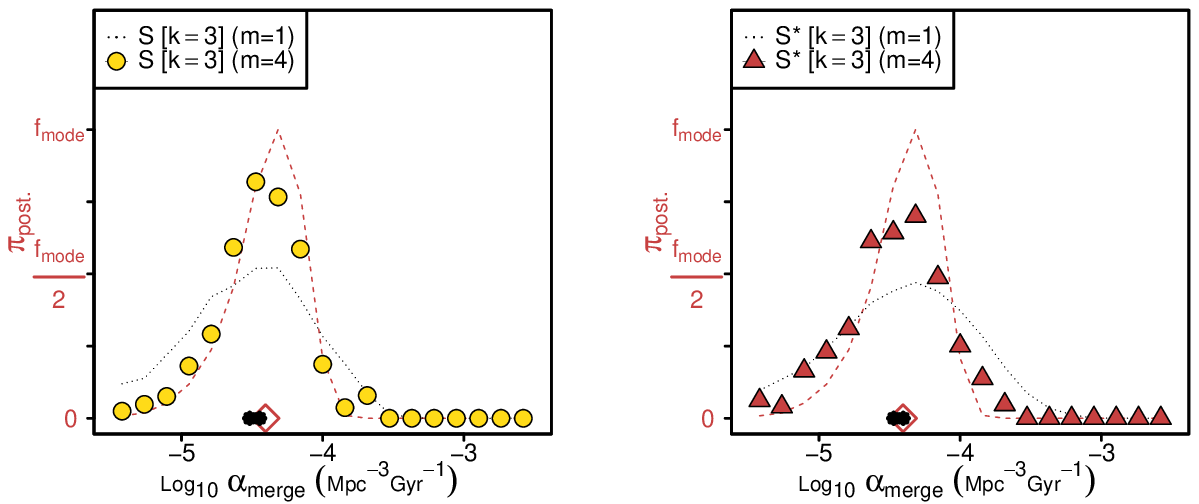}
\caption{Comparison of the marginal posterior density for
  $\alpha_\mathrm{merge}$ in the ``independent evolution'' case of our
  model recovered from
  SMC ABC under the alternative summary statistics, ``na\"ive''
  $\bm{S}: k=3$ and optimised $\bm{S}^\ast : k=3$.  In each case the dotted line
  represents the posterior after one rejection--resampling iteration ($m=1$) of the SMC ABC
  algorithm, the solid datapoints the same after four
  iterations ($m=4$), and the dashed red line the benchmark
  posterior from ``tractable likelihood''-based MCMC.  At the
  bottom of each panel we indicate also the position of the posterior mean
  for this parameter under each of our SMC ABC runs as well as from
  our MCMC
  benchmark (the latter highlighted with a red diamond).\label{summstatsx}}
\end{figure*}

Although $k=3$ proved to be the optimal binning scheme for the set of ``na\"ive''
summary statistics examined above one cannot simply assume this to hold for
the new $\bm{S}^{\ast}$, so we again examine the merits of each
alternative, $k=\{1,3,6,12\}$, following \citet{nun10}.  The results
of this analysis are overlaid against our measurements for the
``na\"ive'' $\bm{S}$ in
Figure \ref{entropy}.   Reassuringly, with the exception of the most limited ($k=1$)
binning scheme these new summary statistics significantly out-perform the
old in the pseudo-RSSE criterion for which they are designed.  Despite our caution $k=3$ does again appear to
consistute the best choice of binning, though $k=6$ is not far behind in accuracy and
may also offer slightly
lower entropy.  In Figure \ref{summstatsx} we compare the marginal
posterior density recovered for the model parameter,
$\alpha_\mathrm{merge}$, following a full run of SMC ABC under
$\bm{S}^\ast : k=3$ against that obtained earlier under $\bm{S} :
k=3$.  Interestingly, though our ``na\"ive'' summary provides a
visually ``closer'' fit to the shape (especially the width, i.e.,
standard deviation) of the benchmark MCMC density for this parameter, the
optimised statistic
does outperform it with regard to the recovery of the posterior mean (which it
manages within a small tolerance after only a single iteration of the
SMC ABC algorithm).  Hence, given both its ease of implementation and
its demonstrated effectiveness in the present analysis we can
confidently recommend the ``semi-automatic'' scheme of \citet{fea12}
for summary statistic refinement.

\textit{\textbf{A Note on One Alternative}}\ \ As mentioned in the
introduction to this Section the subject of summary statistic selection
for ABC analysis remains an active area of research in the statistical
literature, hence it is worth reviewing here briefly another popular alternative
we have neglected to demonstrate above for the sake of
brevity.   Namely, the ``approximate sufficiency'' algorithm
of \citet{joy08} for iteratively building a master summary statistic from the union
of randomly trialled candidates, with each new addition accepted only
if it offers an improvement in parameter inference exceeding some
threshold.  This algorithm may be of particular interest for SAM-based
studies of galaxy formation given the wide variety of available observational
benchmarks from which summary statistics may be composed (see Footnote
\ref{samstat}), though it has been criticised for a dependence
on the (random) order in which the candidate statistics are tested at
each application (i.e., the stated search procedure is far from
exhaustive).  Note also
that although our above demonstration of the \citet{nun10} procedure
is presented in terms of selecting a
unique summary statistic from four evidently degenerate choices
(i.e., $k=\{1,3,6,12\}$) an optimal union of summary statistics may
also be identified via the minimum entropy/MRSSE criterion, though possibly at
great computational expense if the original set of basis candidates is
large and there are many permutations of interest.

\subsection{SMC ABC Posteriors for Our Stochastic Model of
  Morphological Transformation}\label{smc}
\begin{figure*}
\hspace{-0.15cm}\includegraphics[width=16cm]{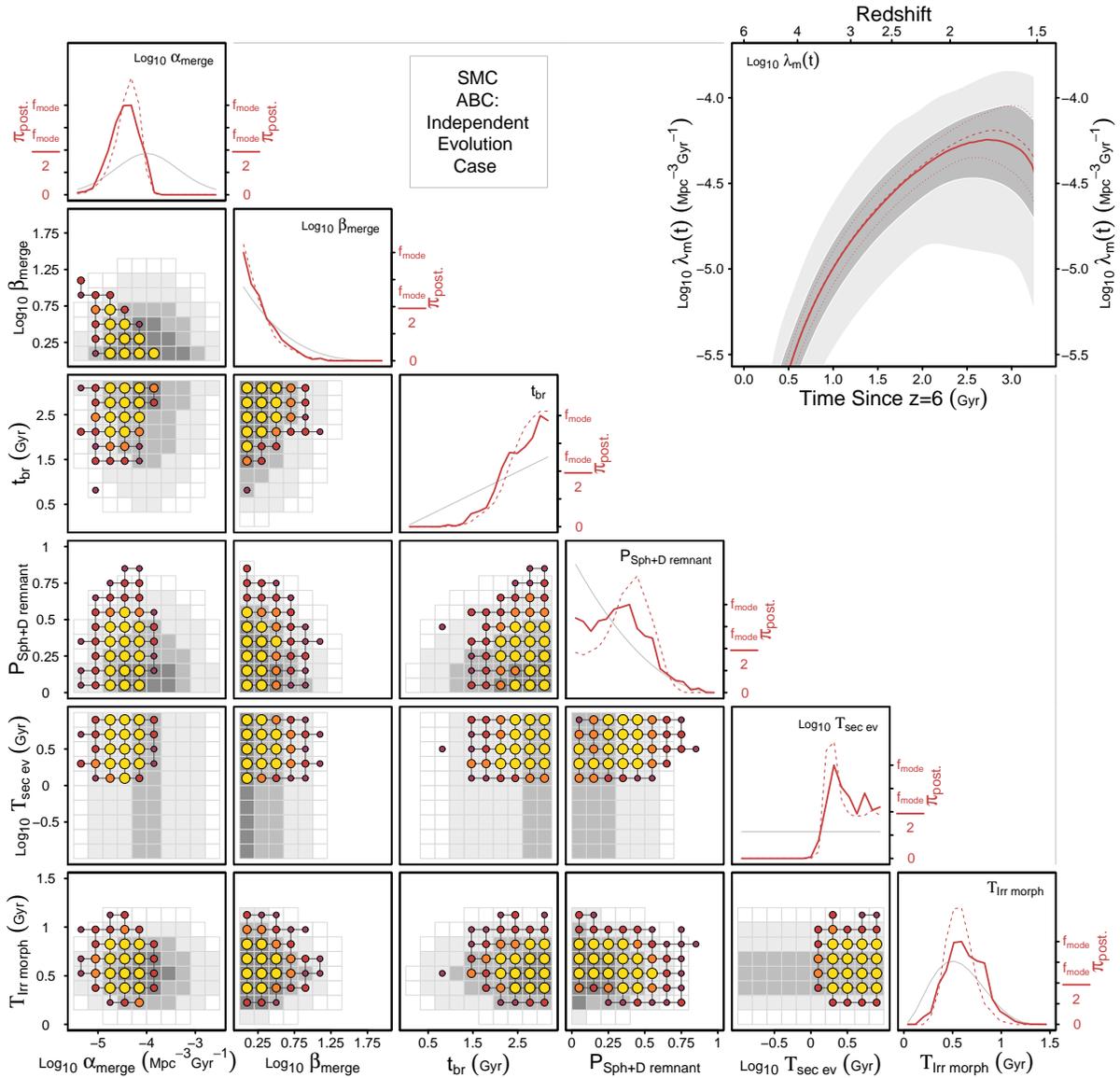}
\caption{SMC ABC posterior probability densities for the key
  parameters of our stochastic model of morphological transformation
  at high redshift
  (in the ``independent evolution'' case).  In each of the main
  diagonal panels we compare the marginal posterior density of a single
  parameter (in red) against its prior (in grey), while in each of
  the off-diagonal panels below we extend this comparison to the joint
  density formed by pairing that parameter against
  one of its peers.  For the latter visualisation we employ
  a lattice of variable-sized points to
  trace the SMC ABC posterior on a
  scale of 1, 2.5, 7.5, and 15 times some appropriate baseline probability density, while grey-shaded tiles map the corresponding prior on
  an identical scale.  In the upper right panel we plot
  the
  (pointwise) 1$\sigma$ and 3$\sigma$ credible intervals and median curve (in dark
  grey, light grey, and red respectively) for the $M_\mathrm{gal}>10^{11}$$M_\odot$ merger rate,
  $\lambda_m(t)$, deriving from our (joint, marginal) posterior densities on
  $\alpha_\mathrm{merge}$, $\beta_\mathrm{merge}$, and
  $t_\mathrm{br}$.  Both here and in the main diagonal panels the MCMC
  (``tractable likelihood''-based) benchmark solution is illustrated
  for comparison via the corresponding dashed (and dotted) lines.\label{abcposteriors}}
\end{figure*}

In Figure \ref{abcposteriors} we present posterior probability
densities for the key parameters of our stochastic model of
morphological transformation at high redshift in the ``independent
evolution'' case, as recovered from SMC ABC using our optimised
summary statistic (cf.\ Section \ref{sumstats} above),
$\bm{S}^\ast:k=3$.  The approximate solution shown here represents the
state of a 10,000 particle population progressed through four
rejection--resampling iterations with an $\alpha=0.75$ rejection rate
and a $c=0.90$ target refreshment rate (cf.\ Section \ref{smcabcalg}).
Comparison against our ``tractable likelihood''-based MCMC
benchmark (for this tractable case of our model)
presented earlier in
Figure \ref{mcmcposteriors} (and overplotted for illustrative purposes here in key panels) highlights the value of the ABC
approach.  That is, without reference to the explicit likelihood function of the
system at hand this simple procedure
of strategic simulation and discrepancy thresholding has nevertheless
produced a most satisfactory approximation to the true posterior,
capturing the key features of each marginal and bivariate joint
density under investigation.  As is expected though
(cf.\ \citealt{csi10}) the ABC posterior does not reproduce exactly the true (``tractable likelihood''-based)
solution here, owing to the inherent gap between ``close'' and ``equal to'' 
in its likelihood approximation---both in the non-zero tolerance for
 the simulated--observed dataset discrepancy required to achieve a
 workable acceptance rate and in the
 fundamental degeneracy of summary statistic matching over full dataset matching
 (cf.\ \citealt{fea12}, for instance).  Moreover, we note that the ABC
 credible intervals so derived (see,
in particular, those for the $M_\mathrm{gal} > 10^{11}$$M_\odot$
merger rate, $\lambda_m(t)$, shown in the top right panel) are for the same reason
noticeably broader than those of the benchmark solution.  Whilst one
would not usually even consider an ABC approach if the likelihood were
tractable it is reassuring for those occasions of interest in which it is not to
verify through the above comparison that this
 approximate likelihood scheme can at least perform similarly. 

\begin{figure*}
\hspace{-0.15cm}\includegraphics[width=16cm]{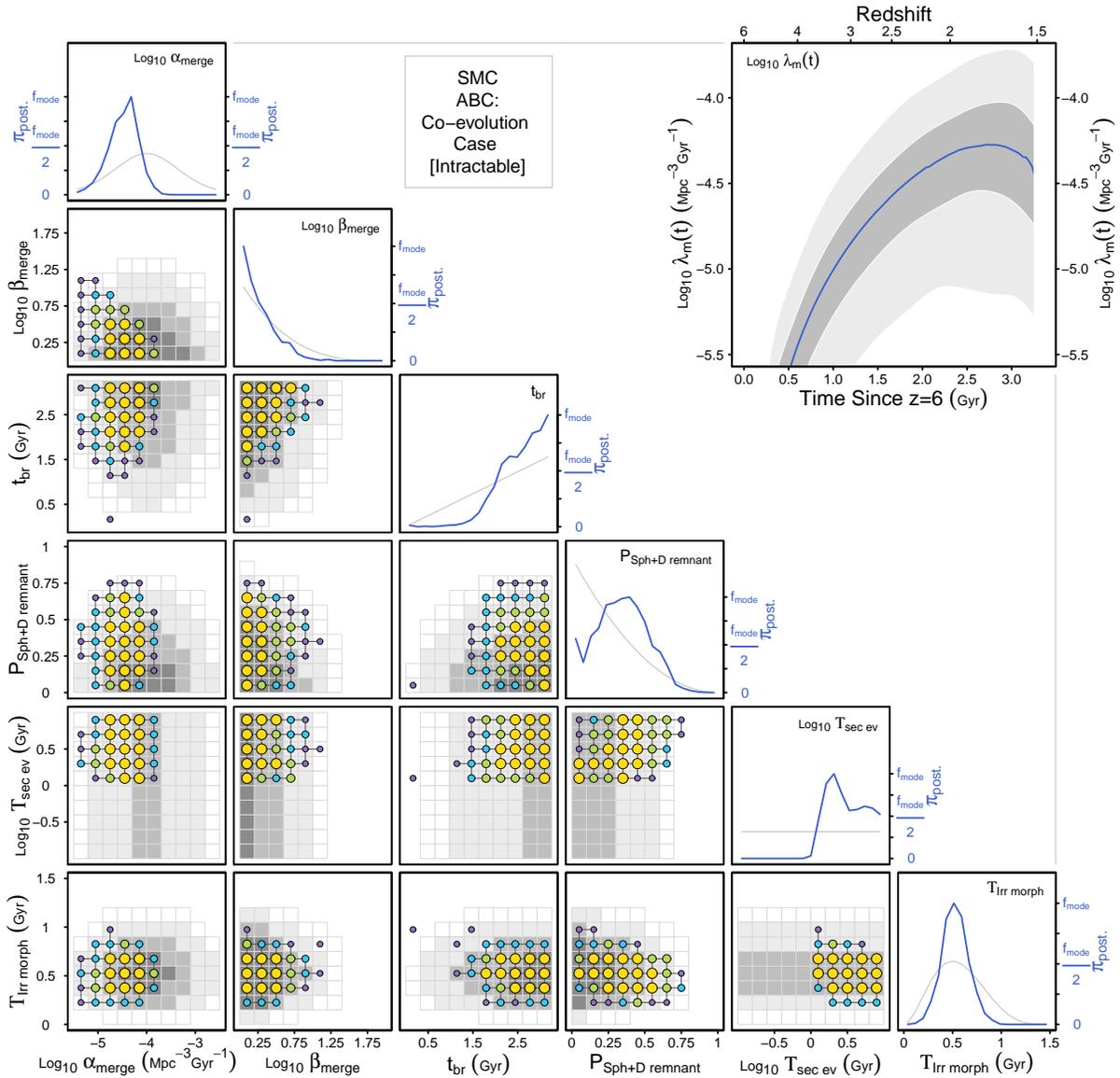}
\caption{SMC ABC posterior probability densities for the key
  parameters of our stochastic model of morphological transformation
  at high redshift
  (in the ``co-evolution'' case).  In each of the main
  diagonal panels we compare the marginal posterior density of a single
  parameter (in blue) against its prior (in grey), while in each of
  the off-diagonal panels below we extend this comparison to the joint
  density formed by pairing that parameter against
  one of its peers.  For the latter visualisation we employ
  a lattice of variable-sized points to
  trace the SMC ABC posterior on a
  scale of 1, 2.5, 7.5, and 15 times some appropriate baseline probability density, while grey-shaded tiles map the corresponding prior on
  an identical scale.  In the upper right panel we plot
  the
  (pointwise) 1$\sigma$ and 3$\sigma$ credible intervals and median curve (in dark
  grey, light grey, and blue respectively) for the $M_\mathrm{gal}>10^{11}$$M_\odot$ merger rate,
  $\lambda_m(t)$, deriving from our (joint, marginal) posterior densities on
  $\alpha_\mathrm{merge}$, $\beta_\mathrm{merge}$, and $t_\mathrm{br}$.\label{coevposteriors}}
\end{figure*}

In Figure \ref{coevposteriors} we present the SMC ABC posteriors
recovered from the
``co-evolution'' case of our model for which indeed the likelihood function is
(by construction for this example) thoroughly intractable---and thus the standard toolbox of
(``tractable likelihood''-based) MCMC
simulation entirely inaccessible.  As described in Section
\ref{stochastic} this intractability is induced simply by coupling the birth times of galaxies in close
associations in a manner consistent with that observed in the
\citet{del07} SAM; leaving all other details of the model unchanged.
Hence it is perhaps unsurprising that the posteriors for this example
differ only slightly from those presented above, with a modest decrease in confidence regarding the true value of the merger rate
(i.e., the joint, marginal density of $\alpha_\mathrm{merge}$,
$\beta_\mathrm{merge}$, and $t_\mathrm{br}$) and a modest
increase in confidence regarding the merger visibility timescale (from
$\tau_\mathrm{Irr\ morph} \approx 0.63 \pm_{0.18}^{0.20}[1\sigma]$ to
$\tau_\mathrm{Irr\ morph} \approx 0.53 \pm_{0.12}^{0.13}$).
Nevertheless this demonstrated ability of ABC to handle models with intractable
likelihoods, and thus to permit the derivation of robust Bayesian
constraints from arbitrarily ``realistic'' (i.e.,
complex) simulations, offers a wealth of possibilites for
astronomical studies far beyond the present example which cannot
be over-stated.

\section{Astrophysical Results \& Discussion}\label{results}
Having completed our exposition of the ABC algorithm in Section
\ref{method} above we take the opportunity here to explore a
number of
interesting astrophysical results arising from our chosen case study
in morphological transformation at high
redshift.  In Section \ref{majormerger} we compare our SMC ABC-based constraints on
the evolving merger rate in the early Universe against
recent estimates from the literature based on simple close pair and
asymmetric galaxy counts, highlighting the superior
informative power of the former over the latter.  Then in Section
\ref{earlybulge} we discuss our (posterior) preference for merging over secular
evolution as the dominant pathway to early bulge formation in the context of contemporary hydrodynamical and ``semi-empirical'' simulations.

\subsection{The Evolving Merger Rate at the Highest Redshifts}\label{majormerger}
As mentioned in the Introduction to this paper the recent installation
of WFC3 on HST has at last made accessible at high resolution the
rest-frame optical morphologies of massive galaxies at the epoch of peak cosmic star
formation
and AGN activity ($z\sim2$; \citealt{lil96,mad96,oes12,war94}),
opening a unique window into the structural assembly of this first
generation of Hubble sequence analogues \citep{cam11,con11a,szo11a}.
  A particular motivation for our
present
case study in morphological transformation was to
highlight the potential of model-based demographic analysis as a means
to exploit this wealth of new data.  In this Section we thus explicitly demonstrate the
advantages of such an approach (as implemented here via SMC ABC) over
the standard ``one type at a time'' mode of study.

Given the (expected) importance of merging as a driver of both star
formation and morphological transformation under the canonical
hierarchical clustering paradigm \citep{whi78} it is perhaps no great surprise that a total of four separate teams have recently attempted constraint of the $z \gtrsim 1.5$ (major) merger rate based on close pair and/or
asymmetric galaxy counts in deep near-IR imaging.  Namely,
\citet{blu09,blu11} in the GNS field, \citet{man11} in the COSMOS
field (in the sub-region of HST/NICMOS coverage), \citet{law11} in a dedicated HST/WFC3 study
from Cycle 17 sampling multiple fields, and
\citet{wil11} in the UDS field; the first two
exploring the same $M_\mathrm{gal} > 10^{11}$$M_\odot$ regime as in
our paper and the latter two probing much further down the mass
function.

Having identified their target population of impending
mergers (in the case of close pair selection) or recent merger
remnants (in the case of asymmetric galaxy selection) each of these
teams has then proceeded to estimation of the early Universe merger rate in
the following manner.  First, they compute the merger
fraction as the number of mergers detected divided by the total number of galaxies
within the target redshift interval and mass range minus the expected
proportion of false detections (arising, for instance, from chance alignments
along the line-of-sight).\footnote{Following
\citet{con06}, \citet{blu09,blu11} advocate correction of this raw
merger fraction, which they denote $f_m$, to a ``galaxy merger
fraction'', denoted $f_{gm}$, via the relation, $f_{gm} =
\frac{2f_m}{1+f_m}$.  The motivation for this correction is to
identify the ``number of galaxies merging as opposed to the number of
mergers''.  However, we believe this to be a false move in context
of their analysis of the most massive galaxies in the GNS, in which
they identify impending mergers as those $M_\mathrm{gal} >
10^{11}$$M_\odot$ systems host to a close companion ($<30$ kpc) of
similar brightness (i.e., within $\pm$1.5 mag in the observed $H$-band;
corresponding to a lower bound on the mass ratio of $\sim$1:4).  Owing to the steepness
of the galaxy stellar mass function at $z\gtrsim 1.5$ the vast majority of such
companions will almost certainly be sub-$10^{11}$$M_\odot$---indeed in our
sample we find just one close pair in which both are truly high mass galaxies, and \citet{man11} find just two.  Hence, by
``correcting'' to $f_{gm}$ as above one is in fact estimating the merger rate
experienced by both massive galaxies and the (ill-defined) population of less massive
galaxies that will ultimately merge with them, which does not seem a particularly
useful exercise.\label{false}}  Second, they adopt a third-party estimate of the
merger visibility timescale based on N-body/hydrodynamical simulations
of galaxy-galaxy collisions; e.g., \citet{man11} adopt
$\tau_\mathrm{close\ pair}
\sim 0.4\pm 0.2$ Gyr from \citet{lot08}.  And third, they either
estimate the comoving volume density of their target galaxy population
directly from the total count in the survey volume (as in \citealt{man11}), or adopt again a (hopefully
more robust, i.e., less prone to cosmic variance error) third-party
estimate from the literature (as in \citealt{blu09} who take
$\Phi_\mathrm{>10^{11}M_\odot}(z)$ from \citealt{dro05}).  The merger
rate by volume is then computed simply as $f_m \times \Phi / \tau$.

\begin{figure*}
\hspace{-0.85cm}\includegraphics[width=11.25cm]{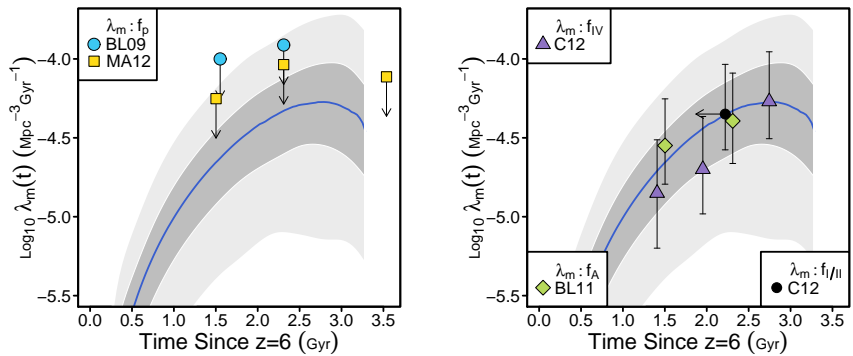}
\caption{The evolving merger rate over the first $\sim$3 Gyr of
  massive galaxy formation ($z\sim6$ to $z\sim1.5$).  Here $\lambda_m(t)$ represents the rate by volume of
  major (\textit{viz.} morphology-changing) mergers experienced by
  those systems already more massive than $10^{11}$$M_\odot$ prior to
  this accretion event.   As in Figure \ref{coevposteriors} above we plot
  the
  (pointwise) 1$\sigma$ and 3$\sigma$ credible intervals and median
  curve deriving from our SMC ABC analysis of the B11 (CANDELS/EGS)
  morphological mix in dark
  grey, light grey, and blue, respectively.  The close pair
  count-based estimates of BL09 and MA12 (with the former scaled down by a factor of 2 to remove their
  ``correction'' from $f_m$ to $f_{gm}$; cf.\ Footnote
  \ref{false}) are overlaid in the lefthand panel;
  marked here as loose upper bounds
   since their threshold of a 1:4 mass-ratio for
  designation as a ``major'' merger may well be overly generous (e.g.\
  \citealt{hop09a,hop09b} favour 1:3 alternatively).  As a more
  faithful comparison in
  the righthand panel we present asymmetric galaxy count-based estimates derived from
  BL11 (following application of our in-house calibrations for
  $\Phi_{>10^{11}M_\odot}(z)$ and $W$; see Section \ref{stochastic} and Figure
  \ref{birthprocess}).  Also overlaid here are raw Type \textsc{iv}
  count-based estimates from our (C12) visual classifications, along
  with an estimate based on the observed Type \textsc{i} and Type
  \textsc{ii} count in our lowest redshift bin; the arrow on the latter noting the fact
  that the marked position of this datapoint corresponds to the
  limiting case of all these mergers having occurred with only just
  enough time to allow fading of the characteristic post-merger irregular features
  prior to observation.
\label{mergerrates}}
\end{figure*}

In the lefthand panel of Figure \ref{mergerrates} we compare
the resulting estimates of $\lambda_m(t)$ from the close pair studies
of \citet{blu09} and \citet{man11} (denoted BL09 and MA12 here,
respectively) against the credible intervals derived from our SMC ABC
analysis of the B11 CANDELS/EGS morphological mix (in the
``co-evolution'' case of our model).  We present the
aforementioned pair count estimates---which are all well outside our
model-based 1$\sigma$ credible interval---as upper bounds on the major
merger rate, since we believe their use of a 1:4 mass ratio
threshold may admit a significant number of ``weak'' accretion events
unlikely to generate a substantial change in morphology; e.g.\
\citet{hop09a,hop09b} favour instead a 1:3 mass ratio for the major/minor
distinction\footnote{As a first-order estimate of the downwards revision
  in these merger rates that would result from switching to a 1:3 mass
  ratio selection one can suppose (perhaps na\"ively) that the masses of close
  pair galaxies correspond to independent random draws from the $z\sim1.5$
  luminosity function; in which case, $\Delta \log
  \lambda_\mathrm{m} \sim -0.15$ to $-0.35$, bringing the
  \citet{blu09} and \citet{man11} determinations broadly into agreement with
  our own.}.  Interestingly, the \citet{man11} results are at least
qualitatively consistent with our posterior inference for the location
of the break epoch at $t_\mathrm{br}\sim2.55$ Gyr (since $z=6$; i.e.,
$z\sim1.75$)---though we note in any case that the plotted datapoints
do carry rather large
uncertainties (of $\sim$88\% at the 1$\sigma$ level; \citealt{man11}), owing in particular to the contribution from
cosmic variance error (and necessarily stellar mass estimation error;
cf.\ \citealt{bra11} and Section \ref{stochastic}) in the determination of $\Phi_\mathrm{>10^{11}M_\odot}(z)$.

As a more faithful comparison---that is, a comparison against rival estimates
based also on post-merger (not pre-merger) observational signatures---we present in the righthand panel of Figure \ref{mergerrates} the merger rate inferred from application of
the $f_m \times \Phi / \tau$ formula to both the asymmetric galaxy
counts of \citet{blu11} (denoted here as BL11) and the raw Type \textsc{iv}
counts of our B11 CANDELS/EGS dataset (denoted C12).  For the
former we adopt $\tau_A \approx 0.6\pm0.3$ Gyr from \citet{con09} (see also
\citealt{lot08}) and for the latter our prior of $\tau_\mathrm{Irr\
  morph}\approx0.55 \pm 0.25$ Gyr; and in both cases we employ our in-house
estimate of $\Phi_{>10^{11}M_\odot}(z)$, which has been calibrated against a large
compilation of recent datapoints from the literature in a manner
accounting for the
presence of systematic
biases in the underlying SED-fit based stellar masses of each
contributing survey (see Section \ref{stochastic} and Figure
\ref{birthprocess}).  Given our definition of $\lambda_m(t)$ as
the rate of mergers experienced by those systems already in excess of $10^{11}$$M_\odot$
at the stated epoch we must allow for the presence of
galaxies only promoted above this mass threshold by the aforesaid
accretion event by
scaling back our raw count-based estimates according to the factor,
$\frac{1}{1+W}$, with $W$ the nuisance parameter introduced in Section \ref{stochastic}.  The error
bars accompanying each datapoint in this righthand panel of Figure
\ref{mergerrates} denote the 1$\sigma$ credible intervals accounting for
the relevant uncertainties in these
estimates (including, of course, those on the estimation of the population
proportion from binomial count data, often mis-handled in astronomical studies; cf.\ \citealt{cam11a}).

Immediately evident from inspection of this righthand panel of Figure
\ref{mergerrates} is the reasonable agreement (well inside the
1$\sigma$ errors)
between our Type \textsc{iv} count-based estimates and those based on the asymmetric
galaxy counts from \citealt{blu11}, confirming a fair degree of
equivalence between the use of visual classification and non-parametric,
quantitative indicators for the selection of high redshift mergers.
Perhaps the most striking impression made by this comparison, however, is the
marked offset between the median curve of our SMC ABC constraint on
$\lambda_m(t)$ from full demographic analysis of the B11 (CANDELS/EGS)
sample and those estimates based only upon the Type \textsc{iv} count
at $t \lesssim 2.5$ Gyr (since $z=6$; i.e., at  $z\gtrsim 2.25$).
 In particular, our median curve here favours a higher merger
 rate, effectively splitting the difference against the higher
 asymmetric galaxy count-based datapoint at $t\sim 1.5$ Gyr (since $z=6$).
 The explanation for this offset lies, of course, in the fact that we fit our
 model not just against the merger fraction but rather the full morphological mix, meaning that the fitted merger
 rate must not only account for the number of ongoing
 mergers at a given epoch but also the observed population of
 evolved systems (i.e., ellipticals and bulge-dominated disks) which
 must have (or in the latter case, \textit{will very likely} have; see Section
 \ref{earlybulge} below) undergone a merger in their recent past.  As an indication of the contribution of Type
 \textsc{i} and Type \textsc{ii} counts\footnote{In fact, we
   downweight the count of Type \textsc{ii} galaxies in this calculation by
   $(1-P_\mathrm{Sph+D\ remnant})$ to acknowledge the (minor) role of secular
   evolution in early bulge formation, as discussed further in Section \ref{earlybulge} below.} to our fit of $\lambda_m(t)$
 we have also marked in Figure \ref{mergerrates} a pseudo-$\lambda_m$
 datapoint estimated by treating these evolved galaxy
 types as ongoing mergers observed at $t_\mathrm{obs}-\tau_\mathrm{Irr\
   morph}$.  The pseudo-$\lambda_m$ datapoint thus computed confirms
 that the past rate of merging was very likely to have been higher
 than that indicated by our raw Type \textsc{iv} counts.  Our counts of Type \textsc{i} and
 \textsc{ii} galaxies also contribute a valuable source of data for
 constraining
 $\tau_\mathrm{Irr\ morph}$, which through our ABC analysis we verify is indeed probably close to our
 prior expectation (i.e., our prior mean of $\tau_\mathrm{Irr\
   morph}\approx0.55\pm0.25$ falls well within the 1$\sigma$ credible interval of
 our posterior, $\tau_\mathrm{Irr\ morph} \approx 0.53
 \pm_{0.12}^{0.13}$).  The associated increase in confidence regarding
 the true value of $\tau_\mathrm{Irr\ morph}$ contributes
 significantly to the reduced width of our SMC ABC credible
 intervals on $\lambda_m(t)$ (based on the full demographics) relative to those based only on our Type \textsc{iv} counts.

\subsection{The Dominant Role of Merging over Secular Evolution for
  Early Bulge Formation}\label{earlybulge}
Another interesting feature of the ABC-based model constraints derived in
this paper concerns the relative dominance of merging over secular
evolution as the favoured mechanism responsible for building up the
first generation of massive bulges in early-type disks.  In particular, the posterior
probability density of our $P_\mathrm{Sph+D\ remnant}$ parameter
favours production of a Type \textsc{ii} system in $\approx 33 \pm 17$\% $[1\sigma]$ of
mergers at these high redshifts---where the gas-rich nature of the
progenitors has previously been argued as conducive to disk survival
and/or rapid reformation around a central spheroid on the basis of hydrodynamical simulations
(e.g.\ \citealt{rob06,hop09a}; but see \citealt{bou11} regarding the
difficulties of reproducing such merger outcomes in models with a realistically
cold, turbulent interstellar medium).  Interestingly, \citet{hop09b}
have also argued for Type \textsc{i} suppression in gas-rich mergers as a
solution to the inconsistency between conventional SAMs and the
observed bulge and disk demographics in the local Universe.  The
posterior density on the secular evolution timescale for our model,
$\tau_\mathrm{sec\ ev} \approx 2.8 \pm_{1.2}^{3.8}$ $[1\sigma]$ Gyr,
on the other hand, renders this rival pathway to bulge formation rather
unlikely for $z\gtrsim1.5$ disks.  Indeed given the above posterior
means one can confirm via repeated simulation from our model that on average only $\sim7$\% of Type
\textsc{ii} systems detected at these redshifts will have been formed via secular
evolution.  This result is consistent with the
recent observations of \citet{gen10} and \citet{hop11} from
hydrodynamic simulations in which wind-driven feedback appears to destroy all
but the most massive clumps in less time than required for their inwards
migration under dynamical friction.

\section{Conclusions}
In this paper we have demonstrated the potential of ``Approximate
Bayesian Computation'' for astronomical model analysis through a detailed
case study in the morphological transformation of high redshift
galaxies.  In the process we have derived tight constraints on the
evolving merger rate in the early Universe, and revealed the relative
dominance of merging over secular evolution for bulge formation
at these epochs, through an ABC-based
examination of the full population demographics of an $M_\mathrm{gal} >
10^{11}$$M_\odot$, $1.5 < z < 3$ sample from the B11 CANDELS/EGS
dataset.  More importantly though our exposition of the contemporary
``Sequential Monte Carlo'' implementation of ABC as well as two
modern approaches to summary statistic selection will hopefully
guide and inspire further astronomical applications of this powerful
statistical technique.

\section*{Acknowledgments}
\texttt{[1]} This work is based on observations taken by the CANDELS Multi-Cycle
Treasury Program with the NASA/ESA HST, which is operated by the
Association of Universities for Research in Astronomy, Inc., under
NASA contract NAS5-26555.

\noindent\texttt{[2]} This study makes use of data from AEGIS, a multiwavelength sky survey conducted with the Chandra, GALEX, Hubble, Keck, CFHT, MMT, Subaru, Palomar, Spitzer, VLA, and other telescopes and supported in part by the NSF, NASA, and the STFC.

\noindent\texttt{[3]} E.C.\ is grateful for financial support from the Australian Research Council.

\clearpage
\appendix
\section{Likelihood Computation for Our Stochastic Model of
  Morphological Transformation in the ``Independent Evolution'' Case}
In this Appendix we derive the likelihood function, $P(\bm{y}|\bm{\theta})$, of the
observed data---i.e., the (HST WFC3/IR) $H$-band demographics for our
sample of 126 galaxies at $1.5 < z < 3$ and
$M_\mathrm{gal} > 10^{11}$$M_\odot$ selected from the B11
CANDELS/EGS dataset---given a particular set of input
parameters,
$\bm{\theta}=\{\alpha_\mathrm{merge},\beta_\mathrm{merge},t_\mathrm{br},P_\mathrm{Sph+D\
  remnant},\tau_\mathrm{sec\ ev},\tau_\mathrm{Irr\ morph}\}$, under the
``independent evolution'' case of our stochastic model for high redshift
morphological transformation.  To this end we must first compute
generic expressions for the probability densities of the time of birth and the time of last
major merger given an arbitrary redshift of observation.  Integration
 of the conditional transition probabilites of
the permitted evolutionary pathways to a given morphology as a
function of the above returns the likelihood of observing that type at
a particular redshift, and via the independence assumption allows a
simple formulation of the complete likelihood function.

\textit{\textbf{Probability Density for the Time of Birth}}\ \ The ``birth'' of
galaxies in our model (i.e., the promotion, via star formation or
merging, of new systems to the
top end of the $1.5 < z < 6$ stellar mass function) is characterised as a non-homogeneous
Poisson process of rate,
\[
\lambda_b(t) = 10^K t^{\gamma},
\]
in units of Mpc$^{-3}$Gyr$^{-1}$ with
the origin of the time variable set to $z=6$.  Both $K$ and $\gamma$ are
treated here as nuisance parameters with $f_{K,\gamma} \sim
\mathcal{N}_\mathrm{Trunc.}([-4.1,0.65]^{\prime},[0.06^2,0.1^2;\rho=0.05];0 < \gamma <
1)$. The waiting time
distribution for galaxy births under the above-specified Poisson
process\footnote{We note for reference that \citet{gla11} provide a brief review of the
  fundamentals of Poisson processes in their recent paper on the Drake
  equation.} is, of course, exponential in $\Lambda_b$-space, where
\[
\Lambda_b(t) =
\int_{0}^{t} \lambda_b(t) dt = \frac{10^K t^{\gamma+1}}{\gamma+1}
\mathrm{.}
\]
Given that each galaxy experiences (by definition) only a single
birth the $\Lambda_b$-epoch of this event represents a unique draw
from the Uniform distribution on $[0,\Lambda_b(t_\mathrm{obs})]$ with
$t_\mathrm{obs}$ the cosmological time since $z=6$ at the observed
redshift, $z_i$.  Transformation of variables back to the time domain
specifies a probability density for the \textit{time} of birth,
$t_\mathrm{birth}$, on $[0,t_\mathrm{obs}]$ of
\[
f_b(t_\mathrm{birth})dt_\mathrm{birth}
=
\frac{\gamma+1}{t_\mathrm{obs}^{\gamma+1}}t_\mathrm{birth}^{\gamma}dt_\mathrm{birth}
\mathrm{.}
\]

\textit{\textbf{Probability Density for the Time of the Last Major
    Merger}}\ \ As for the case  of the birth function
examined above, merging is treated under our stochastic model as a non-homogeneous Poisson process,
with a variable rate by volume (in Mpc$^{-3}$Gyr$^{-1}$) set by the input parameters,
$\alpha_\mathrm{merge}$, $
\beta_\mathrm{merge}$, and $t_\mathrm{br}$, of
\[
\lambda_m(t) =\left\{
\begin{array}{ll}
\frac{\alpha_\mathrm{merge}\beta_\mathrm{merge}}{{t_\mathrm{br}}^2}t^2
& \mbox{for } 0 \le t
\le t_\mathrm{br},\\
\left\{\begin{array}{l} \alpha_\mathrm{merge}\beta_\mathrm{merge}- \\
  \frac{(t-t_\mathrm{br})\alpha_\mathrm{merge}(\beta_\mathrm{merge}-1)}{t_{1.5}-t_\mathrm{br}}
  \end{array} \right. & \mbox{for } t_\mathrm{br} < t \le
t_{1.5}.
\end{array}
\right.
\]
Here $t_{1.5}$ is used to denote the cosmological time between $z=6$ and $z=1.5$ (the
lower bound of our sample).  The number
of mergers experienced by an individual galaxy prior to observation for a given birth time is thus Poisson
distributed with rate,
\[
\Gamma_m^{\ast} = \int_{t_\mathrm{birth}}^{t_\mathrm{obs}}\frac{\lambda_m(t)}{\Lambda_b(t)}
dt = \Gamma_m(t_\mathrm{obs}) - \Gamma_m(t_\mathrm{birth}),
\]
where $\Gamma_m(t) =$
\[
\left\{
\begin{array}{ll}
\frac{\alpha_\mathrm{merge}\beta_\mathrm{merge}(\gamma+1)}{K
  t_\mathrm{br}^2 (2-\gamma)}t^{2-\gamma}  & \mbox{for } 0 \le t
\le t_\mathrm{br},\\
\left\{
\begin{array}{l} \frac{\alpha_\mathrm{merge}\beta_\mathrm{merge}(\gamma+1)}{K
  (2-\gamma)}t_\mathrm{br}^{-\gamma}+ \\
\frac{\alpha_\mathrm{merge}\beta_\mathrm{merge}(\gamma+1)}{K\gamma}(t_\mathrm{br}^{-\gamma}-t^{-\gamma})-
\\
\frac{\alpha_\mathrm{merge}(\beta_\mathrm{merge}-1)}{t_{1.5}-t_\mathrm{br}}\times
\\ \frac{\gamma+1}{K}(\frac{t^{1-\gamma}}{1-\gamma} +
\frac{t_\mathrm{br}t^{-\gamma}}{\gamma} -
\frac{t_\mathrm{br}^{1-\gamma}}{\gamma(1-\gamma)}) \end{array}\right. & \mbox{for } t_\mathrm{br} < t \le
t_{1.5}.
\end{array}
\right.
\]
That is, the probability of a galaxy experiencing $k$ mergers
between its epoch of birth and its epoch of observation is
\[
P(N_m = k|t_\mathrm{birth}) = \frac{
(\Gamma_m^{\ast})^k e^{-\Gamma_m^{\ast}}}{k!}\mathrm{,}
\]
with two cases of particular interest being $P(N_m = 0|t_\mathrm{birth}) =
e^{-\Gamma_m^{\ast}}$ and $P(N_m > 0|t_\mathrm{birth}) =
1-e^{-\Gamma_m^{\ast}}$.  Moreover, for the latter the (non-zero)
$N_m = k$ mergers will be distributed uniformly in $\Gamma_m$-space on $[\Gamma_m(t_\mathrm{birth}),\Gamma_m(t_\mathrm{obs})]$.  The corresponding
probability density of $\Gamma_m$-epoch of the most recent merger is then simply that of the $k$-th order statistic,
\[
f_{\Gamma_m,(k)} (\Gamma_m) d\Gamma_m= \frac{k (\Gamma_m-\Gamma_m(t_\mathrm{birth}))^{k-1}}{(\Gamma_m^{\ast})^k}
d\Gamma_m\mathrm{,}
\]
(remembering that $\Gamma_m^{\ast} =
\Gamma_m(t_\mathrm{obs})-\Gamma_m(t_\mathrm{birth})$).  Summation
over all possible (non-zero) merger counts, $k = 1,\ldots,\infty$,
weighted by their respective probabilites, $P(N_m=k)$, returns the
overall probability density of the most recent merger $\Gamma_m$-epoch as
\[
f_{\Gamma_m^r} (\Gamma_m)d\Gamma_m = \frac{ \sum_{k=1}^{\infty}
  f_{\Gamma_m,(k)} (\Gamma_m) P(N_m=k|t_\mathrm{birth}) }{P(N_m>0|t_\mathrm{birth})} d\Gamma_m
\]
\[
= \frac{ \sum_{k=1}^{\infty} \frac{k
(\Gamma_m-\Gamma_m(t_\mathrm{birth}))^{k-1}}{(\Gamma_m^{\ast})^k} \frac{(\Gamma_m^{\ast})^k
e^{-\Gamma_m^{\ast}}}{k!}  }{1-e^{-\Gamma_m^{\ast}}} d \Gamma_m
\]\[
 =
\frac{e^{\Gamma_m-\Gamma_m(t_\mathrm{birth})}}{e^{\Gamma_m^{\ast}}-1} d \Gamma_m \mathrm{.}
\]
Once again transformation of variables delivers the form of this density
back in the time domain. Namely,
\[
f_{t_m^r} (t_m)dt_m = \frac{e^{\Gamma_m(t_m)-\Gamma_m(t_\mathrm{birth})}}{e^{\Gamma_m^\ast}-1}\frac{\lambda_m(t_m)}{\Lambda_b(t_m)}dt_m.
\]

\textit{\textbf{Likelihood of a Type III Late-Type Disk}}\ \ As
described in Section \ref{stochastic} galaxies in our model may be born
as either Type \textsc{iii} disks or Type \textsc{iv} ongoing
mergers, with the probability of the latter set by
\[
P(C_i(t_\mathrm{birth})=\textsc{iv}|t_\mathrm{birth},\bm{\theta},\bm{\Theta})
= \frac{W
    \lambda_m(t_\mathrm{birth})}{\lambda_b(t_\mathrm{birth})}.
\]
We note explicitly here the conditional dependence on our suite of nuisance
parameters, $\bm{\Theta} = \{K,\gamma, W\}$; where this final element, which derives
from the shape of the $z\sim1.5$-3 stellar mass function, is taken as $W\approx0.5\pm0.2$, i.e., $f_W \sim
\mathcal{N}_\mathrm{Trunc.}(\mu=0.5,\sigma=0.2;W>0)$.  Since no transformation
process permits a transition (back) to Type \textsc{iii} from outside this
state (i.e., Type \textsc{iii} represents a transient class of the
morphological type Markov chain; see
Figure \ref{pathways}) the corresponding
likelihood is the easiest to derive---it
is merely the probability that a galaxy is born a Type \textsc{iii} disk
and experiences neither merging nor secular evolution prior to observation.

For a given $t_{\mathrm{birth}}$ (with $0 < t_{\mathrm{birth}} <
t_{\mathrm{obs}}$) the
probability of classification $C_i=\textsc{iii}$ under our stochastic model
for morphological transformation is thus
\[
P_{i}(C_i=\textsc{iii}|t_\mathrm{birth},\bm{\theta})
\]
\[
= \iiint_0^{\infty} P(C_i(t_\mathrm{birth})=\textsc{iii}
\cap N_m=0 \cap S=0|t_\mathrm{birth},\bm{\theta},\bm{\Theta})\]
\[ f_{K,\gamma}dKd\gamma f_WdW
\]
\[
=\iiint_0^{\infty}
P(C_i(t_\mathrm{birth})=\textsc{iii}|t_\mathrm{birth},\bm{\theta},\bm{\Theta})P(N_m=0|t_\mathrm{birth},\bm{\theta},\bm{\Theta})
\]
\[
P(S=0|t_\mathrm{birth},\bm{\theta},\bm{\Theta})f_{K,\gamma}dKd\gamma f_WdW\mathrm{.}
\]
Here $P(S=0|t_\mathrm{birth},\bm{\theta},\bm{\Theta})$ represents the null secular
evolution probability, which operates
independently of the null merger probability and the probability of
birth as a Type \textsc{iii} system.  According to the
Gamma-distributed form of the secular evolution timescale under our model,
\[
P(S=0|t_\mathrm{birth},\bm{\theta},\bm{\Theta})
=F_{\mathrm{Gamma}(1+50\tau_\mathrm{sec\ ev},50)}(t_\mathrm{obs}-t_\mathrm{birth}),
\] and, as
noted above,
\[
P(N_m=0|t_\mathrm{birth},\bm{\theta},\bm{\Theta}) =
e^{-\Gamma_m^{\ast}}.\]
Integration of this expression by $t_\mathrm{birth}$ over the
corresponding density, $f_b(t_\mathrm{birth})$, gives the general
likelihood of Type \textsc{iii} morphology for the galaxy at hand,
\[
P_i(C_i=\textsc{iii}|\bm{\theta}) = \int_{0}^{t_\mathrm{obs}}
P_i(C_i=\textsc{iii}|t_\mathrm{birth},\bm{\theta}) f_b dt_\mathrm{birth}.
\]
Analytic solutions for this integral exist in a number of special
cases, such as $\gamma=0$ and $\gamma=1$ (the latter in terms of the
error function), but not to our knowledge for arbitrary, non-integer $\gamma\approx0.65$ as required to fit the observed build up in
number density above $10^{11}$$M_\odot$ (see Figure \ref{birthprocess}).
To evaluate this expression within our MCMC code we thus employ the
efficient numerical technique of Monte Carlo
integration \citep{liu01} at run-time.

\textit{\textbf{Likelihood of a Type I Spheroid}}\ \ In contrast to
the simple case
of the Type \textsc{iii} disk presented above there exist an infinite
 variety of evolutionary pathways
potentially leading to the production of a Type \textsc{i} spheroid under our model.  However, these
pathways may all be considered degenerate with respect to the instance
of one final merger
(possibly that of the galaxy's birth) followed
by fading of all post-merger irregular features and settling of the
merger remnant into a spheroid morphology (see Figure \ref{pathways}).  The
probability of transition through this penultimate step is, of
course, dependent upon the time of the final merger, $t_m$. Following
the derivation above we write down the likelihood of Type \textsc{i}
formation for the $i$-th galaxy with $t_\mathrm{obs}$ corresponding to
$z_i$ as an integral over $t_\mathrm{birth}$ (and now $t_m$ also) as
\[
P_i(C_i=\textsc{i}|\bm{\theta}) = \int_{0}^{t_\mathrm{obs}}
P_i(C_i=\textsc{i}|\bm{\theta},t_\mathrm{birth}) f_b dt_\mathrm{birth}
\]
with
\[
P_i(C_i=\textsc{i}|\bm{\theta},t_\mathrm{birth}) = \iiint_{0}^{\infty}
\lgroup P(N_m > 0 |
t_\mathrm{birth},\bm{\theta},\bm{\Theta})
\]
\[
\int_{t_\mathrm{birth}}^{t_\mathrm{obs}}
P(R=1|t_m,\bm{\theta},\bm{\Theta})  P(E=1|\bm{\theta})  f_{t_m^r}dt_m
\]
\[
+ P(C_i(t_\mathrm{birth})=\textsc{iv}|t_\mathrm{birth},\bm{\theta},\bm{\Theta})P(N_m = 0 |
t_\mathrm{birth},\bm{\theta},\bm{\Theta})
\]
\[
P(R=1|t_\mathrm{birth},\bm{\theta},\bm{\Theta})P(E=1|\bm{\theta})  \rgroup f_{K,\gamma}dKd\gamma f_WdW.
\]
Here $P(R=1|t_m,\bm{\theta},\bm{\Theta}) =
F_{\mathrm{Gamma}(1+100\tau_\mathrm{Irr\
    morph},100)}(t_\mathrm{obs}-t_m)$ represents the probability of the post-merger irregular
features fading to reveal the final remnant morphology prior to
observation, and $P(E=1|\bm{\theta}) = 1-P_\mathrm{Sph+D\ remnant}$ represents the probability that the final remnant emerges a
Type \textsc{i} pure spheroid rather than a Type
\textsc{ii} spheroid-plus-disk.  The first half of this equation
corresponds to the case of Type \textsc{i} classification at
$t_\mathrm{obs}$ for a galaxy which has experienced at least one merger
since birth, while the second
corresponds to the case of a sole, primal merger.  Expansion of this expression in
full returns another integral with no simple closed form, so once
again Monte Carlo integration is ultimately required for its evaluation.

\textit{\textbf{Likelihood of a Type IV Ongoing Merger}}\ \ The
likelihood of observing a Type \textsc{iv} ongoing merger is easily derived from the
case of the Type \textsc{i} spheroid above with the trivial changes required
being a complementation of the probability of the post-merger
irregular features fading and removal of the binomial merger
remnant type probability.

\textit{\textbf{Likelihood of a Type II Spheroid-plus-Disk}}\ \ As
illustrated in Figure \ref{pathways} there are in fact two non-degenerate pathways to
formation of a Type \textsc{ii} spheroid-plus-disk system under our
model for high redshift morphological transformation, corresponding to
merging or secular evolution alternately.  So, in principle,
writing down the likelihood of this type, $P_i(C_i=\textsc{ii}|\bm{\theta})$,
should be the most difficult of all.  However, having derived
likelihoods for  each of the other morphological types one may simply evaluate this
case as
\[
P_i(C_i=\textsc{ii}|\bm{\theta}) = 1 - P_i(C_i=\textsc{i}|\bm{\theta})
- P_i(C_i=\textsc{iii}|\bm{\theta})
\]
\[
- P_i(C_i=\textsc{iv}|\bm{\theta}) \mathrm{.}
\]
Such an approach of course requires accurate evaluation of the
likelihoods for all three alternative
morphologies.  Testament to the robustness of our Monte Carlo
integration approach is the fact that upon setting $P_\mathrm{Sph+D\
  remnant}=0.5$ and $\tau_\mathrm{sec\ ev}=100$ Gyr, for which
theoretically, $P_i(C_i=\textsc{i}|\bm{\theta})=P_i(C_i=\textsc{ii}|\bm{\theta})$, the
computational agreement of these likelihoods (for a single galaxy)
evaluated in \texttt{R} typically extends to at least the 4th
significant figure (with just $n_\mathrm{mc} \sim 1000$).

\textit{\textbf{Likelihood of the Full Dataset}}\ \ Recovering the likelihood of
the full observational dataset, $P(\bm{y},\bm{\theta})$ with
$\bm{y}=\{ \bm{y}_i: (C_i,z_i)\}$ ($i=1,\ldots,N_\mathrm{gal}$), in the ``independent evolution'' case is then
simply a matter of taking the product of likelihoods for each
individual galaxy (as even neighbouring galaxies evolve uncoupled in this
scenario).  Thus,
\[
P(\bm{y}|\bm{\theta}) = \prod_{i=1}^{N_\mathrm{gal}}P_i(C_i|\bm{\theta})\mathrm{.}
\]
Of course, when the assumption of independence is relaxed (in order
to build a more physically realistic model) in the
manner of the
``co-evolution'' case considered in Section \ref{coevsection} this full dataset
likelihood function is no longer so readily tractable, and ABC methods
thus become
 essential for reconstructing the posterior probability densities of the key input parameters.

\textit{\textbf{Markov Chain Monte Carlo Simulation}}\ \ Using the
above expressions for $P(\bm{y}|\bm{\theta})$ one may easily generate the
benchmark (``tractable likelihood'') posteriors shown in Figure \ref{mcmcposteriors}
via the (standard) random walk MCMC algorithm.  That is, from the
current state, $\bm{\theta}_i$, propose a new state,
$\bm{\theta}_{i+1} = \bm{\theta}_i + \bm{\delta}$, by sampling
$\bm{\delta}$ from some zero mean distribution, and accept this proposed state with probability
\[
\mathrm{max}(\frac{\pi(\bm{\theta}_{i+1})P(\bm{y}|\bm{\theta}_{i+1})}{\pi(\bm{\theta}_{i})P(\bm{y}|\bm{\theta}_{i})},1).
\]
Here we employ the symmetric multivariate Normal distribution,
$\mathcal{N}(\bm{0},\bm{\Sigma})$, for $\bm{\delta}$ with $\bm{\Sigma}=\bm{\Sigma}_\mathrm{prior}/5$ chosen (by trial-and-error) to
produce, on average, an $\sim$40\% acceptance rate for
$\bm{\theta}_{i+1}$.  Running two separate threads of \texttt{R} on a 2.7 GHz dual core (4
GB ram), 13 inch Macbook Pro
laptop we were able to verify satisfactory convergence of this chain after
completing a target of 100,000 MCMC steps (with a 1,000 step burn-in
period on each thread) in a little less than 48 hours.

\label{lastpage}
\end{document}